\documentclass[fleqn,10pt,twocolumn]{wlscirep}
\usepackage[utf8]{inputenc}
\usepackage[T1]{fontenc}
\usepackage{bm}
\usepackage{hyperref}
\usepackage{multirow}
\usepackage{caption} 
\usepackage{graphicx}
\usepackage{booktabs}
\usepackage{tabularx} 
\usepackage{array}  
\usepackage{booktabs} 
\usepackage{graphicx}
\usepackage{wrapfig}
\usepackage{hanging}
\usepackage{cleveref}
\usepackage{amsmath}
\usepackage{tipa}

\title{Emotion-Aware Design: Modulating Valence, Arousal, and Dominance in Communication via Design}

\author[1]{Shixiong Cao}
\author[1*]{Nan Cao}
\affil[1]{Intelligent Big Data Visualization Lab, Tongji University, Shanghai, China}

\affil[*]{e-mail: nan.cao@tongji.edu.cn}

\begin{abstract}
In an era of emotionally saturated digital media and information overload, effective communication demands more than clarity and accuracy—it requires emotional awareness. This review introduces the paradigm of emotion-aware design, a framework grounded in the valence–arousal–dominance (VAD) model of affect, which systematically examines how emotional modulation shapes comprehension, memory, and behavior. Drawing on insights from psychology, neuroscience, communication, and design, we show that emotional responses significantly influence how information is perceived, retained, and shared. We further propose a multimodal design space—encompassing text, visuals, audio, and interaction—that enables strategic regulation of emotional dimensions to enhance communication efficacy. By linking emotional dynamics to cognitive outcomes and practical design strategies, this review offers both a conceptual foundation and an applied roadmap for designing emotionally resonant communication across domains such as education, health, media, and public discourse.

\end{abstract}
\begin{document}

\flushbottom
\maketitle

\thispagestyle{empty}





\section{Introduction}
In an age of information overload and emotionally saturated digital media, effective communication requires more than factual accuracy—it requires emotional awareness. Emotions play a central role in how individuals perceive, process, and respond to information. They guide attention, shape memory, influence judgment, and drive behaviors such as sharing and engagement, making them fundamental to the dynamics of communication.    

This review examines the emerging paradigm of emotion-aware design—a design-driven approach to understanding and modulating the emotional dynamics of communication. Grounded in the valence–arousal–dominance (VAD) model from affective science, we explore how targeted emotional modulation can enhance message comprehension, memory, and dissemination.

Central to this review is the principle that information design inherently modulates emotional responses. Multimodal content (text, visuals, audio, interaction) actively shapes valence, arousal, and dominance (VAD), directly influencing information processing and behavioral outcomes. As AI-generated content, social media, and immersive technologies become pervasive, emotion-aware design emerges as both a transformative tool and an urgent ethical imperative. When ethically deployed, it fosters empathy, trust, and engagement; when unconstrained, it risks amplifying misinformation, polarization, and psychological burden.

\begin{figure}[!htb]
\setlength{\intextsep}{10pt plus 2pt minus 2pt}
    \centering
    \includegraphics[width=8.5cm]{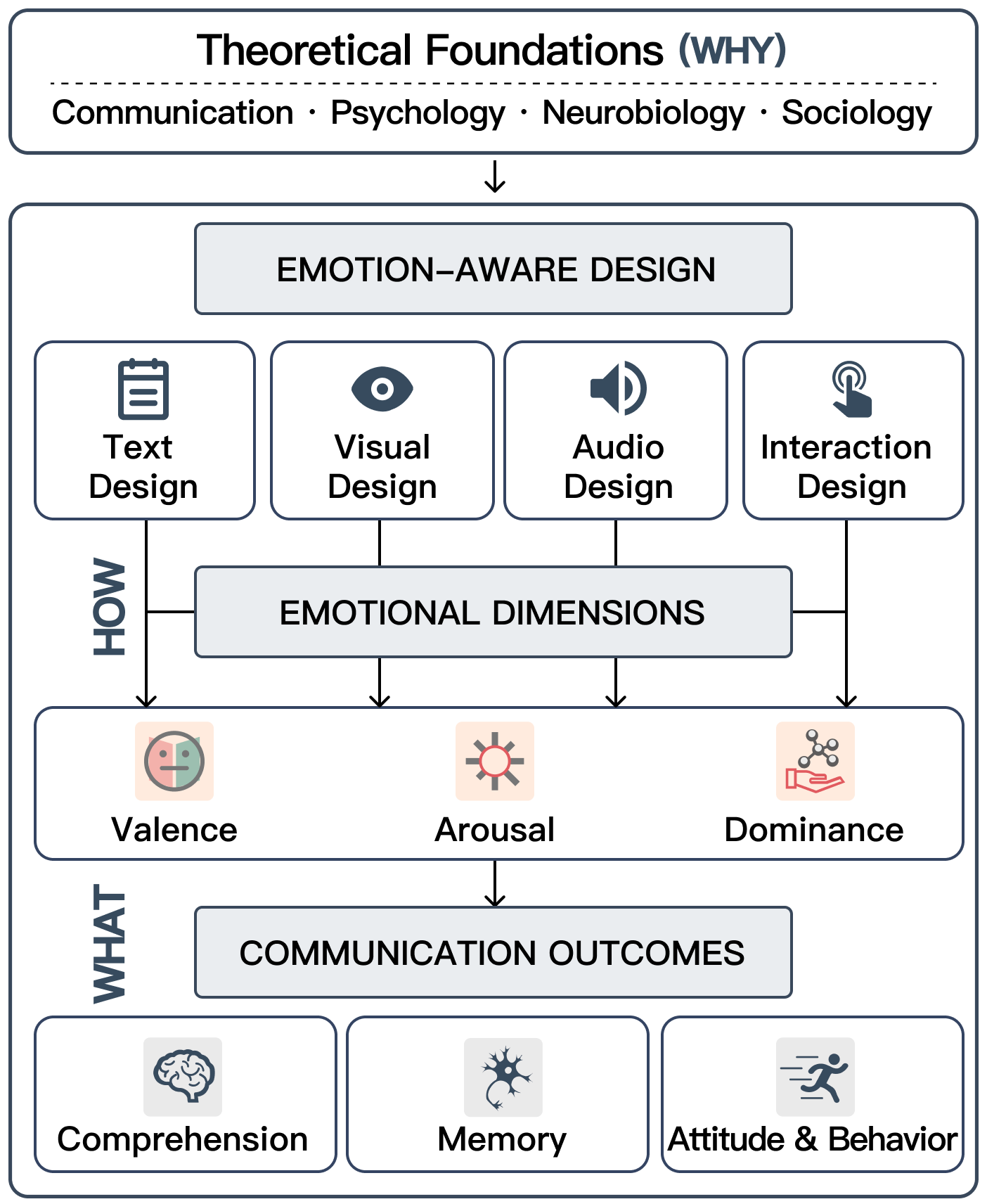}
    \caption{Modulating emotion in communication via design}
   \vspace{-8mm}
\label{fig:why-Fig1}
\end{figure}

To navigate these challenges and opportunities, we synthesize insights from psychology, neuroscience, communication theory, and design research. Building on this foundation, we propose a multimodal framework for emotion-aware communication, illustrating how distinct design modalities (i.e., text, visuals, audio, and interaction) can be strategically aligned with specific emotional targets across the valence–arousal–dominance (VAD) space.

This review is organized around three guiding questions(FIG.~\ref{fig:why-Fig1}): (1) Why do emotions matter in information communication? (2) What communication factors are affected by emotions? and (3) How can design be used to regulate emotion and enhance communicative impact? By bridging theoretical insights with design strategies, we outline a roadmap for applying emotion-aware principles across diverse domains—including education, healthcare, media, public discourse, and digital interaction.


\section{Emotions Models} \label{sec:emotion_models}
Researchers have developed various models to explore both the dimensional and compositional aspects of emotion, establishing a robust theoretical foundation for emotion studies. These models are typically classified into two categories: discrete models \cite{plutchik1980general, izard1991psychology, ekman1992argument} and continuous models \cite{russell1980circumplex, mehrabian1974approach, osgood1952nature}.

\textbf{Discrete Models} conceptualize emotions as distinct categories, though theorists differ in the number and nature of these basic emotions. Plutchik \cite{plutchik1980general} introduced the Wheel of Emotions, which identifies eight core emotions (e.g., joy, fear, anger, sadness) and highlights their combinatorial properties and intensity variations—illustrated by how anxiety may emerge from a mix of fear and anger. Izard \cite{izard1991psychology} expanded this framework by proposing ten fundamental emotions, including interest, shame, and guilt, and emphasizing their physiological bases and corresponding facial expressions. In contrast, Ekman \cite{ekman1992argument} identified six universal emotions—anger, fear, happiness, sadness, disgust, and surprise—that exhibit robust cross-cultural consistency and underpin research on emotion recognition and cultural comparisons.


\begin{figure*}[!htb]
\setlength{\intextsep}{10pt plus 2pt minus 2pt}
    \centering
    \includegraphics[width=18cm]{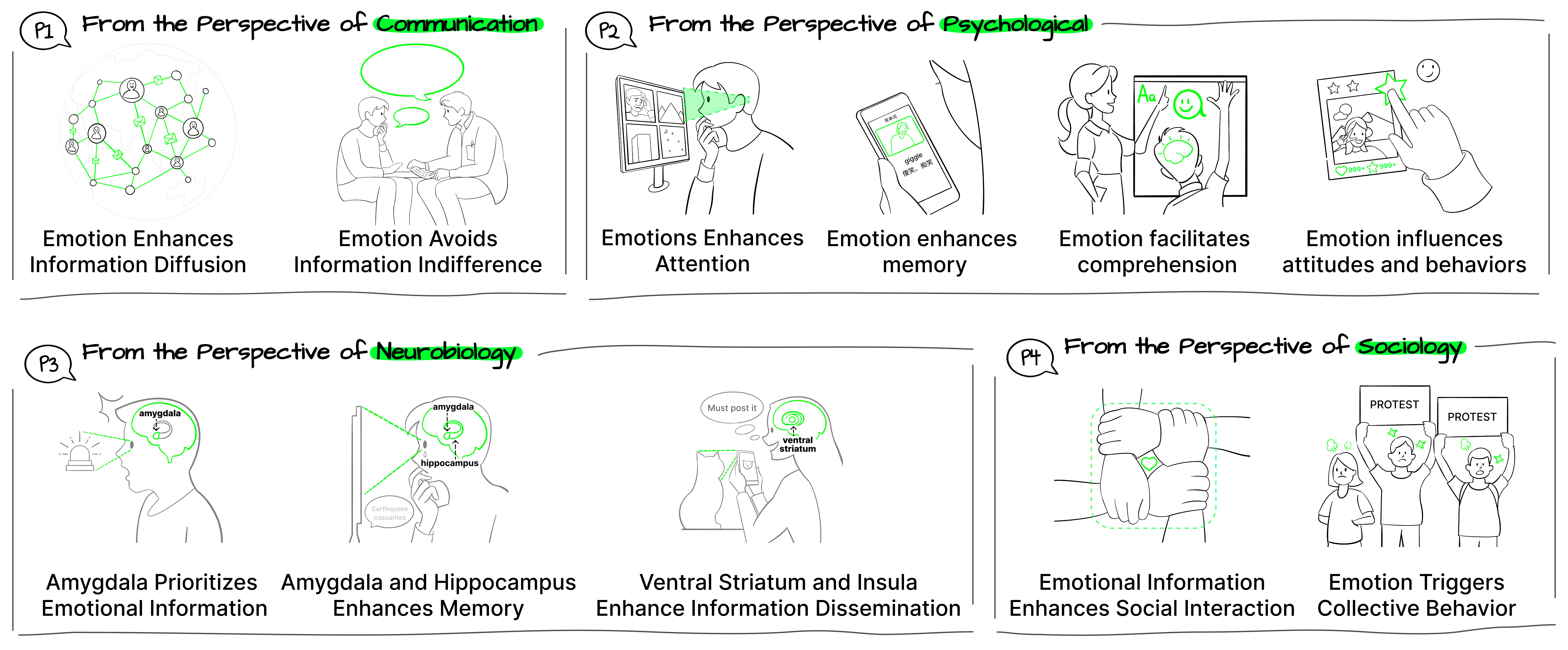}
    \caption{This figure illustrates the role of emotions in information communication from four perspectives: communication (P1), psychological (P2), neurobiology (P3),  and sociology (P4).}
   \vspace{-2mm}
\label{fig:why-Fig2}
\end{figure*}

\textbf{Continuous Models} view emotions as dynamic processes characterized by continuous dimensions. Russell \cite{russell1980circumplex} introduced the circumplex model of affect, which characterized emotional states along two dimensions:  valence, referring to the degree of positivity or negativity of an emotional state, and arousal, denoting the level of physiological and psychological activation. For example, excitement is high in both valence and arousal, whereas depression scores low on both dimensions. Building on this, the PAD model \cite{mehrabian1974approach} introduces a third dimension—dominance—which captures the extent to which an emotion involves a sense of control or power over one’s environment. For example, anger is linked to high dominance, while fear is associated with low dominance. Similarly, Osgood’s model \cite{osgood1952nature} describes emotions through evaluation (valence), activation (arousal), and potency~(dominance).


Although emotion models vary in their frameworks and dimensions, valence, arousal, and dominance (V-A-D) consistently emerge as fundamental descriptors of emotional experience. By modulating key emotional attributes, these dimensions profoundly affect how information is understood, remembered, and disseminated, making them central to the study of information communication.

\section{Why Emotion Matters in Communication?} \label{sec:why} 
Neurobiological studies reveal that the brain processes emotional information differently from neutral stimuli, leading to distinct communication effects \cite{Zheng2017amygdala, alexander2021neuroscience, kensinger2020retrieval}. Research psychology and sociology consistently highlight emotion's crucial role in communication (FIG.~\ref{fig:why-Fig2}). It shapes how information is perceived \cite{vuilleumier2005how}, shared \cite{ecker2022psychological}, and acted upon \cite{xie2024using, smith2024polarization}, underscoring its pivotal influence on communication processes.


\subsection{Emotional Processing in the Brain}
Neurobiological evidence shows that emotional information is processed quickly and encoded through key regions, such as the amygdala \cite{Gothard2020} and hippocampus \cite{alexander2021neuroscience, Zheng2017amygdala}. The amygdala prioritizes emotional stimuli, triggering rapid affective responses before conscious thought \cite{phelps2005contributions}. It works through a subcortical pathway, enabling immediate reactions to threats or rewards \cite{dejong2024state, Gothard2020}. This fast response is crucial for threat detection and social interactions \cite{Gothard2020, phelps2005contributions}.

The amygdala and hippocampus collaborate to enhance emotional memory \cite{Zheng2017amygdala}, with emotional arousal strengthening memories for both positive and negative events \cite{mcgaugh2015consolidating, Gothard2020}. Positive emotions promote dopamine release, enhancing memory consolidation \cite{alexander2021neuroscience}, while negative emotions intensify memory vividness \cite{kensinger2020retrieval}. 

Information spreading is also driven by emotion and involves key brain regions, including the ventral striatum \cite{haber2010reward}, insula \cite{craig2009you}, prefrontal cortex (PFC) \cite{Scholz2017neural}, and empathy-related areas \cite{penagos2022mirror, nejati2023role}. The ventral striatum activates when processing positive emotions, increasing the likelihood of sharing content, which explains why heartwarming and anger-inducing stories spread quickly \cite{haber2010reward, Baek2017value}. The PFC modulates this process by evaluating socio-emotional cues and guiding decisions about sharing \cite{Scholz2017neural}. Empathy-related regions, such as the medial prefrontal cortex (mPFC) and temporoparietal junction (TPJ), enhance emotional resonance, facilitating connection with the content \cite{penagos2022mirror, nejati2023role}. The insula amplifies emotional salience by assigning greater affective weight to emotional stimuli, making them more perceptually prominent \cite{craig2009you}. Meanwhile, the anterior cingulate cortex (ACC) accelerates emotional processing, especially in response to high-arousal emotions like fear and anger, working with the amygdala and ventromedial PFC to enhance the spread of emotional content \cite{etkin2011emotional}.



\subsection{From the Perspective of Cognition}
Emotion significantly affects how we pay attention to and process information. Emotionally significant content naturally captures more attention than neutral stimuli, particularly high-arousal emotions \cite{sussman2013emotional, zsido2020count}. Negative emotions tend to keep attention focused longer \cite{todd2012affect}, while positive emotions broaden attention, enabling the processing of more information \cite{madan2019positive}.

Emotions also enhance memory retention \cite{tyng2017influences}. Content that elicits strong emotional responses—whether positive or negative—is remembered more vividly and accurately \cite{madan2019positive}. Negative emotions sharpen focus on critical details, improving recall precision \cite{todd2020emotional}, while positive emotions enhance the integration of broader concepts, making it easier to form connections between ideas \cite{madan2019positive}.

Furthermore, emotion helps in comprehension by reducing cognitive load \cite{megalakaki2019effects}. Emotionally charged content is processed more efficiently because it activates multiple cognitive channels, making it easier to understand and remember \cite{tyng2017influences}. This is especially evident in high-arousal emotions like fear or excitement, which direct attention to key elements of the content \cite{lane2015memory}.

Especially in today’s media-saturated environment, emotionally infused content engages audiences more deeply, making it more memorable and impactful \cite{stieglitz2013emotions, berger2012makes}. Emotional narratives, especially those that evoke personal connections or social relevance, foster greater attention and involvement than neutral or purely factual presentations \cite{richardson2020engagement, jheng2025strategies}.

\subsection{From the Perspective of Sociology}

From a sociological perspective, emotion shapes social behaviors by influencing attitudes and driving actions\cite{liu2025self, xie2024using, ecker2022psychological}. Studies show that emotional content enhances the resonance of information, prompting users to share content, make purchases, or engage in other actions \cite{liu2025self, ecker2022psychological, pennycook2021shifting}. Positive emotions usually foster favorable attitudes and reduce resistance \cite{liu2025self, ecker2022psychological}, while negative emotions often trigger immediate, action-oriented responses \cite{xie2024using, ecker2022psychological}.

Emotions also strengthen social connections, particularly in collective settings such as political discourse or social movements \cite{smith2024polarization, agostini2021toward}. Both positive (e.g., joy) and negative (e.g., anger, fear) emotions can deepen social bonds, promote shared identity, and motivate collective action \cite{smith2024polarization, rahrig2025examining}. During crises, emotions like fear or anger can drive rapid group responses \cite{smith2024polarization}, fostering cohesion and mobilizing individuals toward a common goal \cite{agostini2021toward}.

Emotionally charged content is more likely to spread than neutral information, especially in digital and social media environments \cite{brady2017emotion, vuilleumier2005how}. High-arousal emotions like anger, fear, or surprise accelerate content sharing \cite{brady2017emotion}. Notably, negative emotions—such as sadness or anger—tend to drive stronger engagement and faster diffusion, making emotional content more shareable \cite{brady2017emotion, vuilleumier2005how}.

\subsection{Conclusion}
Emotion plays a central role in communication by driving attention, enhancing memory, influencing social interactions, and shaping behaviors. Emotionally charged content spreads more quickly, engages audiences more deeply, and is more memorable. By understanding the dynamics of emotional influence, communicators can craft messages that resonate with audiences, encourage sharing, and motivate action. Next, we review the specific aspects of communication that are influenced by emotion.

\section{What Factors are Affected by Emotions?} \label{sec:what} 



Recent research indicates that emotional expression plays a critical role in shaping users’ comprehension\cite{megalakaki2019effects, mather2011arousal, tyng2017influences}, memory\cite{megalakaki2019effects, tyng2017influences, kensinger2007negative}, and behavioral intention\cite{ferrara2015measuring, stieglitz2013emotions, vosoughi2018spread, dabbous2023influence}. Building on the Valence–Arousal–Dominance (V-A-D) model  \cite{russell1980circumplex, mehrabian1974approach} introduced in Section \ref{sec:emotion_models}, this section examines how each emotional dimension influences key processes in information communication.

\subsection{Comprehension}
Emotions critically influence information comprehension by guiding attentional focus \cite{sussman2013emotional} and modulating cognitive resource~allocation \cite{tyng2017influences}. 

\subsubsection{Valence}
\textit{\underline{Positive emotions}} promote global processing and surface-level comprehension, but may reduce attention to detail. Megalakaki et al. \cite{megalakaki2019effects} found that positive-mood individuals did well on surface tasks (84\% accuracy) but had lower inferential question accuracy, perhaps due to heuristic processing. Similarly, Jiménez-Ortega et al. \cite{jimenez2012emotional}  observed longer reaction times (505 ms) and higher error rates (13.2\%) during semantic judgment tasks, indicating less analytical depth.  

\textit{\underline{Negative emotions}} may impair processing fluency but enhance focus and detail-oriented reasoning. Arfé et al. \cite{arfe2023effects} reported longer first fixation durations when participants in negative emotional states read emotionally charged content, indicating increased attentional engagement. Despite lower overall comprehension, individuals in negative emotions performed better on inferential tasks ($p < .001$) \cite{megalakaki2019effects}.

\textit{\underline{Neutral emotions}} support stable cognitive functioning by promoting balanced resource allocation.  Megalakaki et al. \cite{megalakaki2019effects} found that participants in neutral emotional states performed similarly to the positive group on surface-level tasks and to the negative group on inferential tasks, though their overall performance was slightly lower. The absence of emotional cues may constrain deeper processing and hinder the integration of complex information \cite{earles2016memory}.

In summary, valence influences comprehension in distinct ways. Positive emotions aid integrative understanding but may lower analytical precision. Negative emotions enhance detail - oriented processing but often hurt fluency. Neutral states support cognitive stability, though potentially at the expense of processing depth.

\subsubsection{Arousal} 
\textit{\underline{High arousal emotions}} boost cognitive engagement by focusing attention on emotionally salient information.  
For example, in visual tasks, high arousal increases focus by highlighting targets and filtering out distractions \cite{mather2011arousal}. In speech perception, emotionally charged sounds quickly activate temporal and higher-order brain regions, aiding rapid recognition of affective cues  \cite{bestelmeyer2017effects}.  

\textit{\underline{Low arousal emotions}} reduces attention and limits processing depth. 
low-arousal states weaken perceptual contrast and diminish focus on salient stimuli, thereby impairing comprehension efficiency \cite{mather2011arousal}. The effects of arousal are further shaped by emotional valence: high-arousal negative states heighten sensitivity to threat, whereas low-arousal positive states may suppress analytical processing \cite{bestelmeyer2017effects}. However, in tasks involving emotional distractors, low-arousal negative stimuli have been linked to enhanced task-relevant attention, as indicated by faster reaction times relative to high-arousal negative or positive/neutral stimuli. This suggests that under certain conditions, low-arousal negative states may facilitate more efficient attentional resource allocation \cite{sussman2013emotional}. 

In summary, high arousal enhances attention to emotionally relevant content and supports deeper processing. Low arousal may promote cognitive stability but often limits processing depth, with outcomes influenced by emotional valence.

\subsubsection{Dominance} Emotional dominance influences information processing by modulating cognitive focus and attentional control. 
\textit{\underline{High-dominance emotions}}  (e.g., anger, pride) are marked by a strong sense of control and agency, which supports focused and goal-directed information processing. These emotional states allow individuals to prioritize relevant information and maintain sustained attention \cite{mehrabian1996pleasure}. Neurocognitively, they engage brain regions such as the anterior insula and dorsolateral prefrontal cortex, enhancing attentional control and reducing distractibility \cite{jerram2014neural}.

\textit{\underline{Low-dominance emotions}}  (e.g., fear, shame), by contrast,  are associated with diminished control and heightened vulnerability. They activate brain regions related to hypervigilance, such as the posterior precuneus and amygdala, thereby diverting cognitive resources from goal-directed processing \cite{jerram2014neural}. Consequently, individuals in low-dominance states often struggle to sustain attention and integrate complex information, particularly under high cognitive load \cite{watanabe2015neural}. 

In summary, high-dominance emotions support goal-directed attention and cognitive control, while low-dominance emotions heighten threat sensitivity and may reduce processing efficiency.

\subsection{Memorization}
Emotional states also influence both memory formation and retrieval \cite{megalakaki2019effects}, shaping how information is encoded, stored, and later communicated \cite{mather2011arousal, mcgaugh2015consolidating}.


\subsubsection{Valence} 
\textit{\underline{Positive emotions}} enhance memory coherence and associative processing, supporting both integration and retrieval  \cite{megalakaki2019effects}. Compared to neutral states, individuals in positive emotional states perform better on free recall and associative tasks, such as remembering brand names and contextual information \cite{speer2017reminiscing}. 

\textit{\underline{Negative emotions}} enhance attention to detail \cite{xie2017negative, van2015good}, especially in visual memory, by activating the right fusiform gyrus and amygdala \cite{kensinger2007negative}. Kensinger \& Ford \cite{kensinger2020retrieval} found that such states improve sensory detail recall but increase false memory risk \cite{brainerd2008does, kaplan2016emotion}, as attention narrows to core content while peripheral details are reconstructed during retrieval \cite{kensinger2007negative}. Arfé \cite{arfe2023effects} reported higher recall accuracy for emotionally relevant content in negative ($M = 3.09, SD = 0.77$) than neutral states ($M = 2.48, SD = 0.93$).

\textit{\underline{Neutral emotions}} show limited influence on memory enhancement. Individuals in neutral states tend to allocate fewer cognitive resources to encoding and consolidation \cite{kensinger2020retrieval, megalakaki2019effects}, although their more evenly distributed attention may support consistent performance in certain tasks \cite{reisch2020negative, Srinivasan2010}.

In summary, positive emotions improve associative memory and long-term retention but reduce attention to detail. Negative emotions improve detail-oriented recall but increase susceptibility to false memories. Neutral emotions support cognitive stability but offer limited memory enhancement.

\subsubsection{Arousal} 
\textit{\underline{High arousal emotions}}  (e.g., fear, excitement) enhance long-term memory by directing attention toward emotionally salient content  \cite{mather2011arousal}. Marchewka et al. \cite{marchewka2016arousal} found that high-arousal stimuli had higher recognition rates even after six months. Neurologically, high arousal releases norepinephrine and cortisol, promoting hippocampal plasticity and memory consolidation \cite{mcgaugh2015consolidating}. However, when attention is divided, high arousal may impair memory by narrowing focus to central details while neglecting peripheral information, thereby increasing the likelihood of false memories \cite{mather2011arousal, brainerd2008does}.

\textit{\underline{Low arousal}} emotions reduce attentional bias toward core emotional content, supporting more balanced memory encoding. Marchewka et al. \cite{marchewka2016arousal} found lower false recognition rates for low-arousal stimuli ($p < .05$), suggesting greater mnemonic stability. Sussman et al. \cite{sussman2013emotional} observed improved attentional consistency under low-arousal negative states (e.g., sadness). Arnsten \cite{arnsten2009stress} further noted that higher arousal did not enhance spatial working memory, underscoring the cognitive efficiency of low-arousal states in certain contexts.

In summary, high arousal strengthens memory for emotionally central content but increases the likelihood of distortion, while low arousal supports stable and integrative memory formation. These effects are context-dependent and vary based on task demands.

\subsubsection{Dominance} \textit{\underline{High dominance}} emotions enhance memory by facilitating semantic encoding \cite{perry2001hemispheric} and prioritizing goal-relevant content \cite{mneimne2015beyond}. Perry et al. \cite{perry2001hemispheric} found that high-dominance states engage semantic pathways, while low-dominance emotions activate more perceptual–emotional responses, suggesting distinct neural processes. 
Mneimne et al. \cite{mneimne2015beyond} showed that high-dominance positive words (PVHD) are better recalled in the left hemisphere, whereas high-dominance negative words (NVHD) are less recognized in the right. 

\textit{\underline{Low dominance}} emotions, in contrast,  may enhance memory by increasing threat sensitivity \cite{mneimne2015beyond} and facilitating contextual integration  \cite{arifin2007novel}. Mneimne et al. \cite{mneimne2015beyond} found better recall of low-dominance negative words (NVLD) in the right hemisphere, especially in individuals with high behavioral inhibition. Arifin \& Cheung \cite{arifin2007novel} observed higher recognition for high-dominance scenes (e.g., violence, 82\%) than for low-dominance ones (e.g., fear, 75\%), suggesting that low-dominance emotions aid contextual integration but may be less memorable without urgency.

In summary, high-dominance emotions boost memory via semantic encoding, while low-dominance emotions enhance it through threat sensitivity and contextual integration, reflecting distinct cognitive mechanisms.

\subsection{Attitude and Behavior}
Emotions influence attitudes and behavior, particularly in relation to motivation and communication patterns. 

\subsubsection{Valence} 
\textit{\underline{Positive emotions}} strongly influence user attitudes and behaviors. Stieglitz \& Dang-Xuan \cite{stieglitz2013emotions} found that positive tweets spread faster than neutral ones. Berger \& Milkman \cite{berger2012makes} demonstrated that online content with positive valence is more likely to be shared. Even after controlling for factors such as practical utility and article placement, positive content was 13\% more viral than negative content. Additionally, Cheong et al. \cite{cheong2023synchronized} pointed out that people tend to share positive emotions to gain social recognition and strengthen interpersonal bonds. 

\textit{\underline{Negative emotions}} exert complex influences on information sharing. Van Bavel et al. \cite{bavel2020using} demonstrated that emotions such as fear and anxiety undermine social trust, thereby reducing prosocial sharing during crises. Similarly, Dabbous \& Aoun Barakat \cite{dabbous2023influence} found that negative emotions exacerbate information overload and cognitive biases, facilitating the dissemination of misinformation.


\textit{\underline{Neutral emotions}}, exert minimal influence. Ferrara \& Yang \cite{ferrara2015measuring} found that neutral content on Twitter attracted little engagement, while Cheong et al. \cite{cheong2023synchronized} argued that such emotions lack the motivational force needed for social transmission.

In summary, emotional valence influences information-sharing behavior in distinct ways: positive emotions promote dissemination, negative emotions have context-dependent effects, and neutral emotions exert minimal impact.


\subsubsection{Arousal} 
\textit{\underline{High-arousal emotions}} enhance social media information sharing  \cite{ferrara2015measuring}. Berger \cite{berger2011arousal} found that individuals  in high-arousal states (e.g., amusement) more likely to share than in low-arousal ones (e.g., sadness). Similarly, Stieglitz \& Dang-Xuan \cite{stieglitz2013emotions} observed that political tweets with high-arousal emotions spread faster. However, high arousal can also amplify misinformation. Dabbous \& Aoun Barakat \cite{dabbous2023influence} reported that high-arousal negative emotions reinforce beliefs and cause information overload among false-content sharers. 

\textit{\underline{Low-arousal emotions}} generally inhibit information sharing, though exceptions exist. Stieglitz \& Dang-Xuan \cite{stieglitz2013emotions} found that emotions like sadness lacked the motivational force to drive dissemination.  Similarly, Berger \& Milkman \cite{berger2012makes} reported that New York Times articles featuring low-arousal states were shared less than high-arousal ones.  However, Zhang \& Wang \cite{zhang2023despicable} showed that negative emotions like disgust and anger were strategically deployed in pandemic-era political reporting to heighten attention to intergroup conflict and promote sharing within nationalist crisis narratives.


In summary, high-arousal emotions promote sharing by activating cognitive and physiological systems but may also increase the spread of misinformation. In contrast, low-arousal emotions tend to suppress sharing, though they may encourage it in reflective or socially meaningful contexts.

\subsubsection{Dominance}  
\textit{\underline{High dominance emotions}} (e.g., pride, anger) facilitate communication by enhancing individuals’ sense of control and motivational drive.  As proposed in Mehrabian’s PAD model \cite{mehrabian1996pleasure}, such emotions increase one’s willingness to take action and influence others. Xu et al. \cite{xu2023understanding} found that content expressing high-dominance emotions not only encourages users to share it but also strengthens its persuasive impact.

\textit{\underline{Low dominance emotions}} (e.g., fear, helplessness) inhibit information sharing. As Lebel \cite{lebel2017moving} notes, such states lead to withdrawal and risk avoidance, making individuals less likely to share or endorse content publicly. This may stem from concerns about losing control or facing negative evaluation.  

In summary,  high-dominance emotions facilitate active sharing and persuasive messaging, while low-dominance emotions discourage engagement by lowering agency and increasing social withdrawal.

\section{How to Regulate motions?}\label{sec:how} 




Among various strategies for eliciting emotion, shaping the presentation of information through information design is a key approach to enhancing communication effectiveness. In this section, we examine how information design can be strategically used to regulate emotion and enhance communication. In particular, we propose a design space structured around three core emotional dimensions—valence, arousal, and dominance—to provide both theoretical grounding and practical guidance for improving comprehension, memory, and the persuasive impact of information.

\subsection{Overview of the Design Space}
Emotion plays a central role in information communication design. Previous research has identified eight key design elements that influence users’ emotional responses: images, text, audio, color, layout, navigation, feedback, and rewards \cite{alves2020incorporating, lottridge2011affective}. Based on multimodal processing theory and user experience design principles, these elements can be organized into four core dimensions—textual, visual, audio, and interaction \cite{oviatt2000designing}. This classification enhances both conceptual clarity and practical guidance for affective information design. Textual design encompasses headline framing \cite{kourogi2015identifying, kolev2022foreal}, narrative structure \cite{polya2021temporal}, and narrative content \cite{seo2019process}.  Visual design encompasses color \cite{wilms2018color, jonauskaite2019color}, shape \cite{yan2023case, plass2014emotional}, layout \cite{mai2011rule, makin2012implicit, lu2017investigation}, and imagery \cite{lin2023effect}. Audio design involves tone \cite{schirmer2010mark, weinstein2018you}, sound effects \cite{eerola2012timbre, batterink2016phase}, and music \cite{bottiroli2014cognitive, salimpoor2011anatomically, koelsch2014brain}. Interaction design comprises interaction methods \cite{wodehouse2014exploring}, motion effects \cite{wollner2018slow, lockyer2012affective}, and navigation \cite{sundar2014user}. Together, these dimensions constitute a coherent multimodal framework for affective information design.

\subsection{Textual Design}
\begin{figure*}[hbt!]
\setlength{\intextsep}{10pt plus 2pt minus 2pt}
    \centering
     \includegraphics[width=17.6cm]{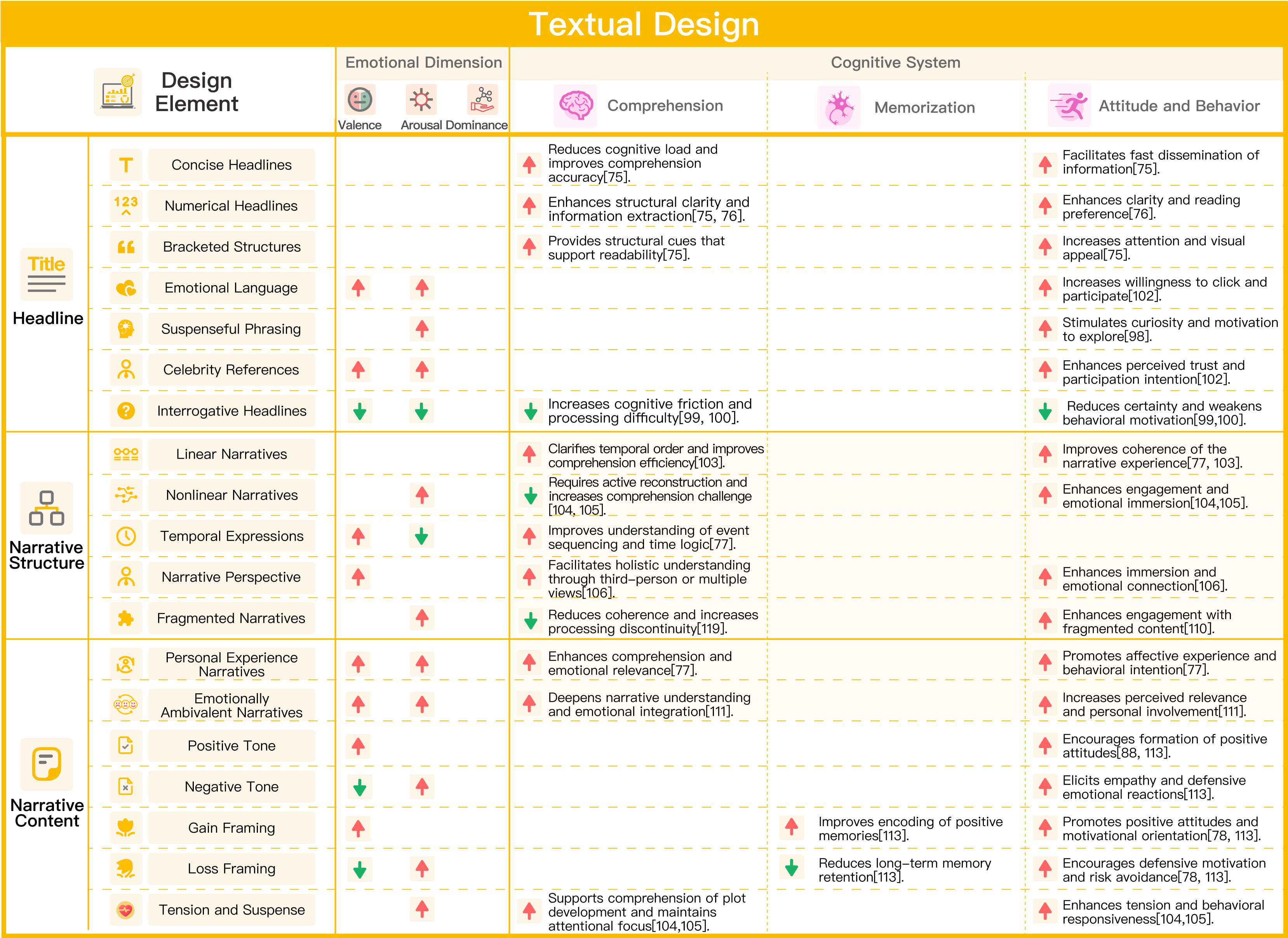}
    \caption{Emotional Dimensions and Cognitive-Behavioral Activation in Textual Design Elements.}
   \vspace{-4mm}
\label{fig:textual_design}
\end{figure*}

Textual design delivers information while shaping readers’ emotional responses through elements such as headlines, narrative structure, and content (FIG.~\ref{fig:textual_design}).

\subsubsection{Headline}

\begin{wrapfigure}{l}{0.06\textwidth}
  \vspace{-11pt} 
    \includegraphics[width=0.07\textwidth, trim=0 0 0 25, clip]{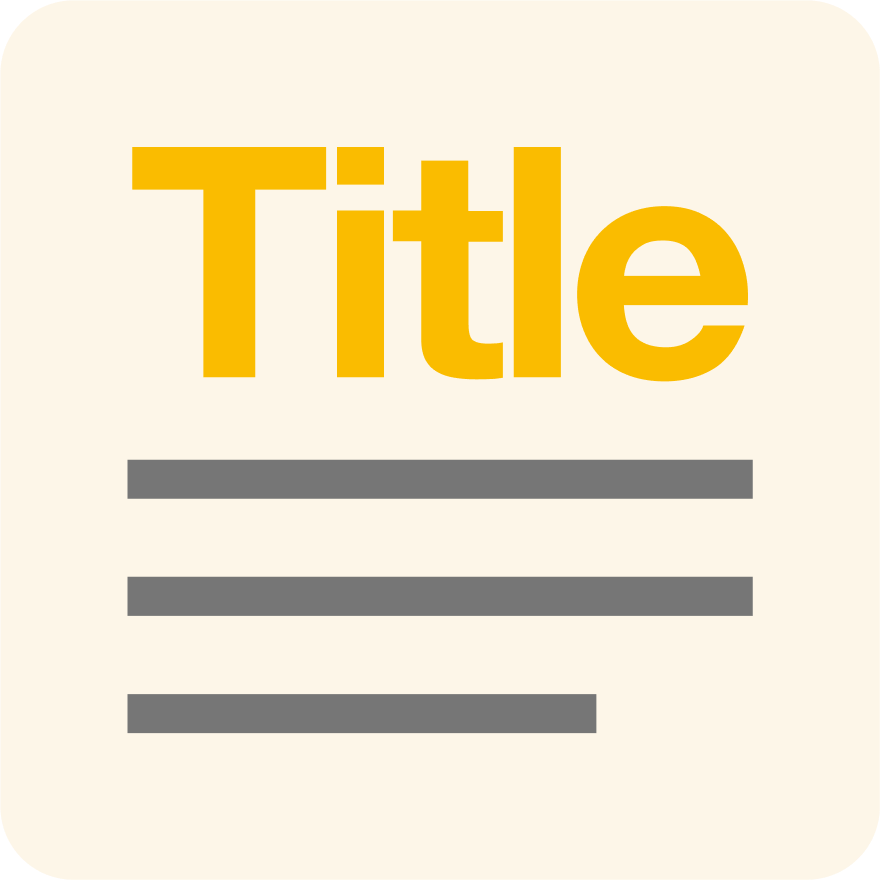}
\end{wrapfigure} 

Headlines function as both summaries and cognitive-emotional cues that guide attention and drive engagement.  
Kourogi et al. \cite{kourogi2015identifying} found that concise headlines reduce cognitive load and improve dissemination, especially on social media. Headlines with bracketed phrases (e.g., “[Expert Advice on XXX]”) or numerical lists (e.g., “5 Techniques to Enhance Writing Skills”) increase readership \cite{kourogi2015identifying}. Kolev \cite{kolev2022foreal} further showed that number-emotion combinations (e.g., “10 Must-Visit Travel Destinations...”) heighten arousal and increase clicks.

Structural elements like forward-referencing \cite{blom2015click} and suspense phrasing \cite{kuiken2017effective} also increase effectiveness.  Blom \& Hansen \cite{blom2015click} found that withholding key details evokes curiosity and encourages clicks, often through demonstratives or signal words (e.g., “this,” “why”) \cite{kuiken2017effective}. Pronouns increased clicks by 25\% ($p<.001$), signal words by 16\% ($p=.002$), while interrogatives reduced engagement by 17\% ($p=.019$), likely due to ambiguity. This reflects how knowledge gaps create a “cognitive itch,” prompting curiosity \cite{golman2015curiosity}, with neuroimaging showing that such knowledge gaps activate dopamine release in the nucleus accumbens \cite{berns2006predictability}.

Emotionally charged language also enhances engagement. Kim et al. \cite{kim2016compete} found affective verbs outperformed nouns, and superlative adverbs exceeded comparatives in click-through rates. Celebrity mentions (e.g., “Amanda Bynes,” $wCTR = 0.171$) further boosted engagement. Overall, emotionally dynamic headlines effectively capture reader attention.

\subsubsection{Narrative Structure}

\begin{wrapfigure}{l}{0.06\textwidth}
  \vspace{-11pt} 
    \includegraphics[width=0.07\textwidth, trim=0 0 0 25, clip]{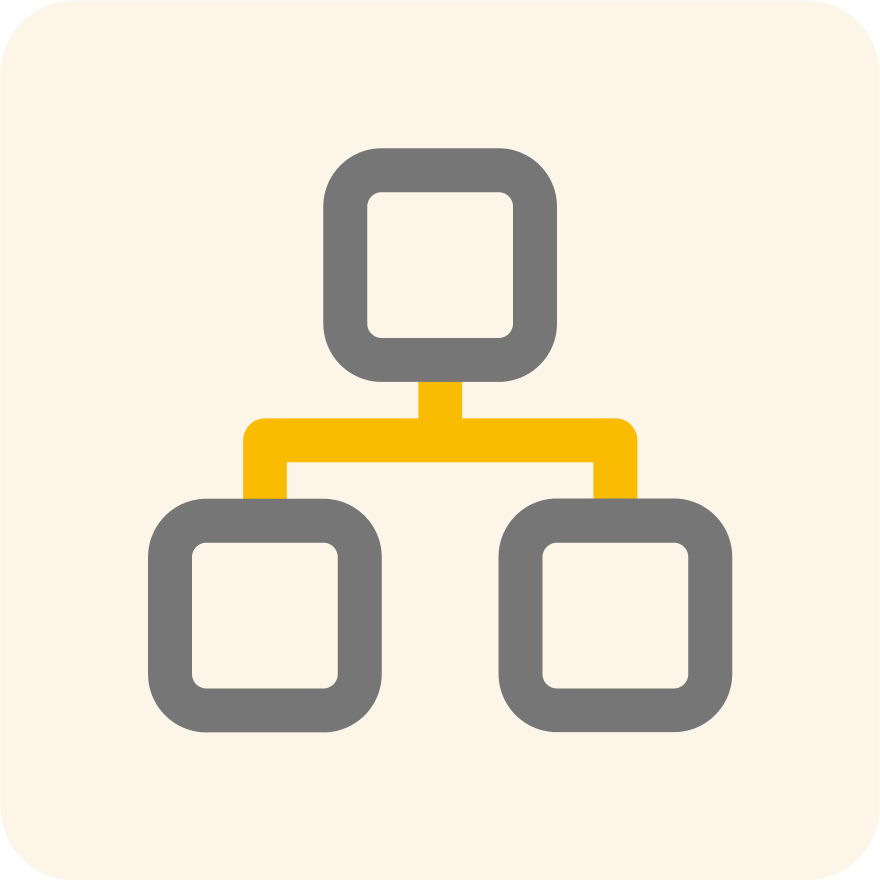}
\end{wrapfigure} 

Narrative structure forms—such as linear versus nonlinear narratives—influence emotional engagement and comprehension by shaping temporal expectations and causal coherence \cite{sanchez2021relative, speer2009reading}. Linear narratives support comprehension by minimizing temporal ambiguity \cite{sanchez2021relative}. Polya \cite{polya2021temporal} found that explicit time markers (e.g., “at 3 p.m.”) negatively correlated with arousal ($r = -0.29$, $p < 0.001$), while state-time descriptors (e.g., “waiting”) positively correlated with valence ($r = 0.18$, $p < .01$). Nonlinear narratives—featuring flashbacks or non-sequential plots—disrupt chronological flow but enhance immersion by encouraging active reconstruction \cite{speer2009reading, zacks2007event}. 

Narrative perspective further affects emotional and cognitive outcomes.  First-person narration increases engagement by providing access to characters’ internal states \cite{libby2011self}, while third-person narration offers broader context, aiding comprehension  and regulation \cite{yarkoni2008neural}. Multi-perspective formats enrich understanding by integrating diverse viewpoints \cite{yarkoni2008neural, kidd2013reading}.

Emerging media formats (e.g., short-form videos, interactive web literature) are reshaping narrative design by challenging traditional structures and perspectives \cite{wang2023narrative}. Short videos emphasize speed over depth, while web narratives use interactivity to boost engagement and flexibility \cite{vijayaraghavan2023m}. These formats raise cognitive load but enable new emotional and informational experiences.

\subsubsection{Narrative Content}

\begin{wrapfigure}{l}{0.06\textwidth}
  \vspace{-11pt} 
       \includegraphics[width=0.07\textwidth, trim=0 0 0 25, clip]{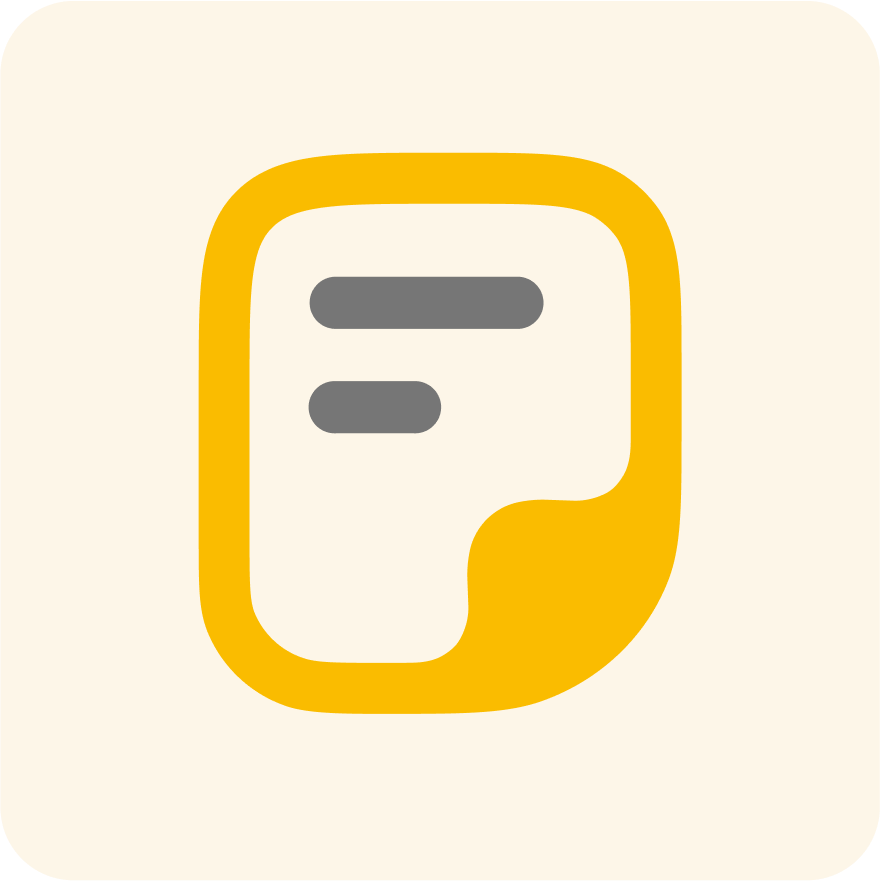}
\end{wrapfigure} 

Narrative content shapes emotional by structuring plot and modulating content tone to guide audience engagement. Personal experience narratives—focused on real-life events and psychological struggles—are especially effective in evoking strong emotions \cite{polya2021temporal}. Reece et al. \cite{reece2023candor} found that emotionally ambivalent stories, combining pain and hope, elicit greater intensity and deeper involvement. Suspense and delayed resolution heighten engagement by gradually revealing conflict, enhancing realism, and promoting message internalization \cite{tchernev2023there}.

Tone, as a key component of narrative content, also influences emotional processing. Seo \& Dillard \cite{seo2019process} found that gain-framed messages (e.g., “This will improve health”) increase positive affect, while loss-framed ones raise anxiety and perceived risk.  Lee \& Potter \cite{lee2020impact} showed that positive words (e.g., “adorable”) triggered approach-oriented affect and enhanced memory (heart rate deceleration: $F(1,50) = 10.17$, $p = 0.002$), whereas negative words (e.g., “painful”) induced defensive responses and distress.
	
\subsection{Visual Design}
\begin{figure*}[hbt!]
\setlength{\intextsep}{10pt plus 2pt minus 2pt}
    \centering
    \includegraphics[width=17.6cm]{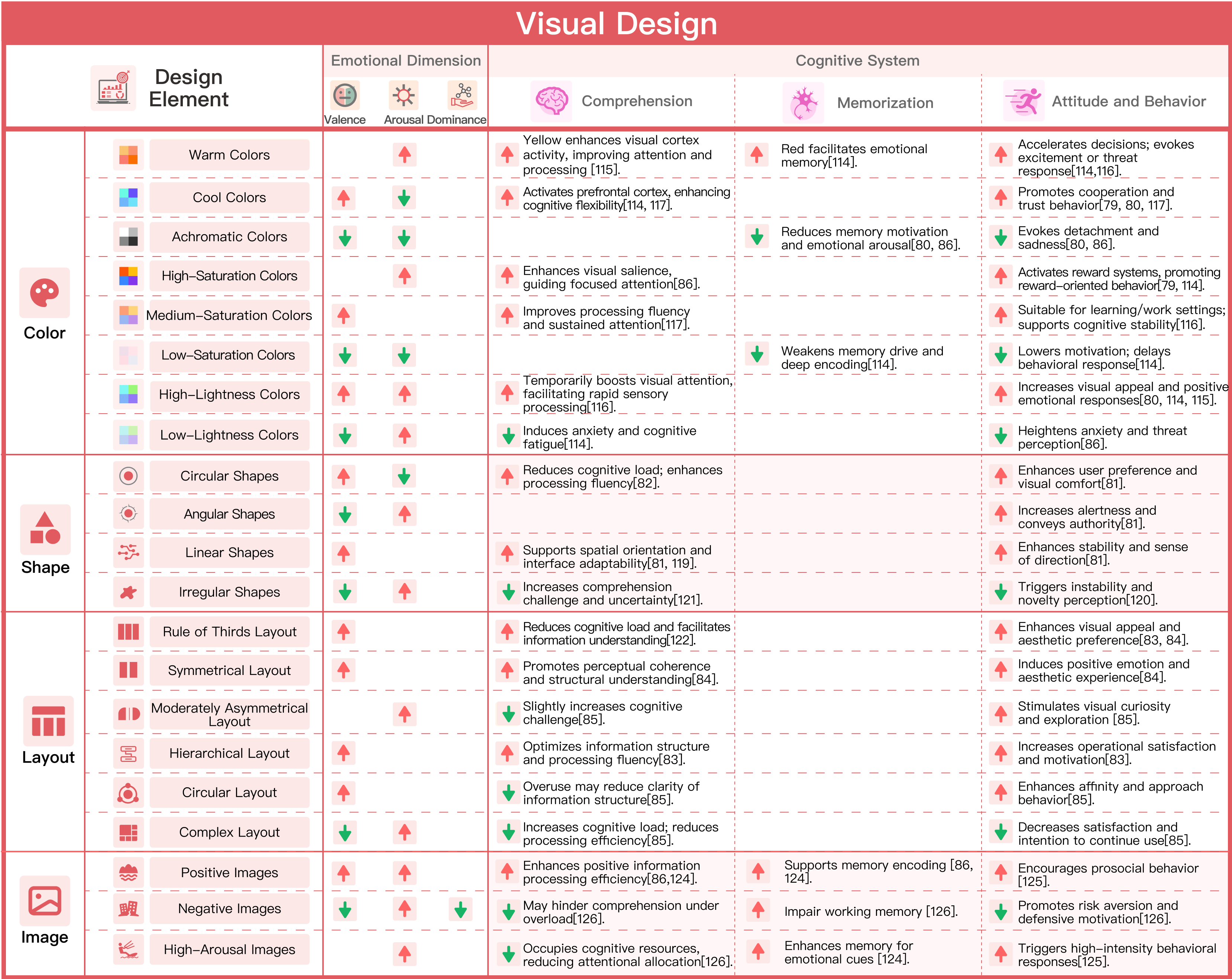}
    \caption{Emotional Dimensions and Cognitive-Behavioral Activation in Visual Design Elements.}
   \vspace{-4mm}
\label{fig:visual_design}
\end{figure*}

Visual design shapes emotional responses through key elements such as color, shape, layout, and imagery (FIG.~\ref{fig:visual_design}). 


\subsubsection{Color}
\begin{wrapfigure}{l}{0.06\textwidth}
  \vspace{-11pt} 
        \includegraphics[width=0.07\textwidth, trim=0 0 0 25, clip]{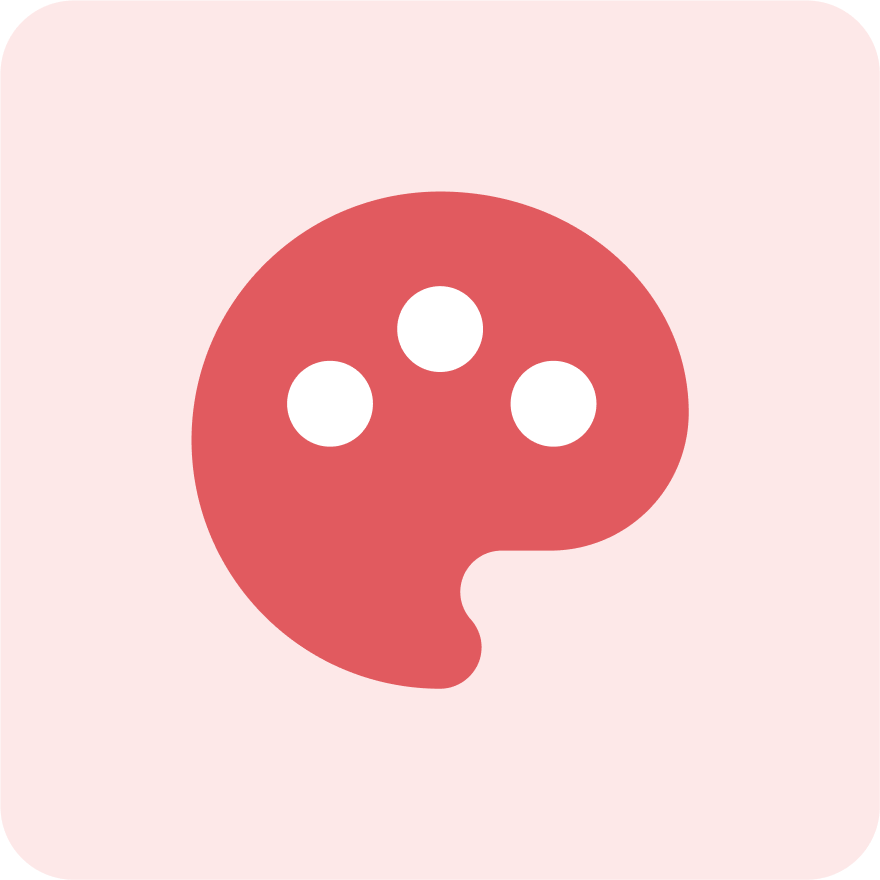}
\end{wrapfigure} 

Color is a fundamental element of visual design, exerting emotional influence through hue, saturation, and brightness.

Hue, which distinguishes colors such as red, blue, and yellow, elicits distinct emotional \cite{weijs2023effects, wilms2018color} and physiological responses \cite{pazda2024colorfulness, jonauskaite2019color}. Warm hues like red are associated with high arousal and moderate valence, activating the sympathetic nervous system—increasing heart rate and skin conductance, while reducing heart rate variability \cite{wilms2018color, weijs2023effects}. These effects enhance excitement and decision speed but may impair creative performance \cite{weijs2023effects, kallabis2024investigating}.  In contrast, cool hues like blue promote high valence and low arousal, fostering calmness, trust, and cooperation, and are linked to reduced amygdala activation and increased prefrontal engagement \cite{jonauskaite2019color, wilms2018color, wang2013interpretable}. Yellow typically induces high valence and moderate arousal, enhancing attention and processing speed through visual cortex stimulation \cite{pazda2024colorfulness, jonauskaite2019color}. Achromatic hues, such as gray, elicit low arousal and valence, suppress motor cortex activity, and evoke detachment or sadness \cite{lin2023effect, jonauskaite2019color}.

Saturation, or color intensity, modulates arousal and attentional focus \cite{jonauskaite2019color}. Highly saturated colors (e.g., vivid red or blue) elevate sympathetic activation and increase physiological arousal \cite{wilms2018color, weijs2023effects}, while also making visual elements more noticeable and emotionally engaging \cite{lin2023effect, pazda2024colorfulness}.  
Low saturation (e.g., gray, pastels) are linked to reduced arousal, slower response times, and diminished motivation \cite{jonauskaite2019color, lin2023effect, weijs2023effects}. 
Medium-saturation colors offer emotional balance, supporting attention and cognitive fluency in environments such as classrooms or digital interfaces \cite{wang2013interpretable, kallabis2024investigating}.

Brightness (perceived luminance) also shapes emotional and cognitive responses. High-brightness colors (e.g., white, light yellow) evoke high valence and moderate arousal, improving perceptual efficiency \cite{jonauskaite2019color, pazda2024colorfulness, weijs2023effects}  and accelerating cognitive processes through visual cortex activation \cite{wilms2018color, kallabis2024investigating}. However, excessive brightness can impair sustained attention and increase fatigue \cite{weijs2023effects}. Low-brightness colors (e.g., navy, dark gray) are associated with negative affect and heightened arousal, activating the amygdala and hypothalamus while reducing prefrontal activity—patterns linked to anxiety and threat perception \cite{weijs2023effects, lin2023effect}.

\subsubsection{Shape}
\begin{wrapfigure}{l}{0.06\textwidth}
  \vspace{-11pt} 
        \includegraphics[width=0.07\textwidth, trim=0 0 0 25, clip]{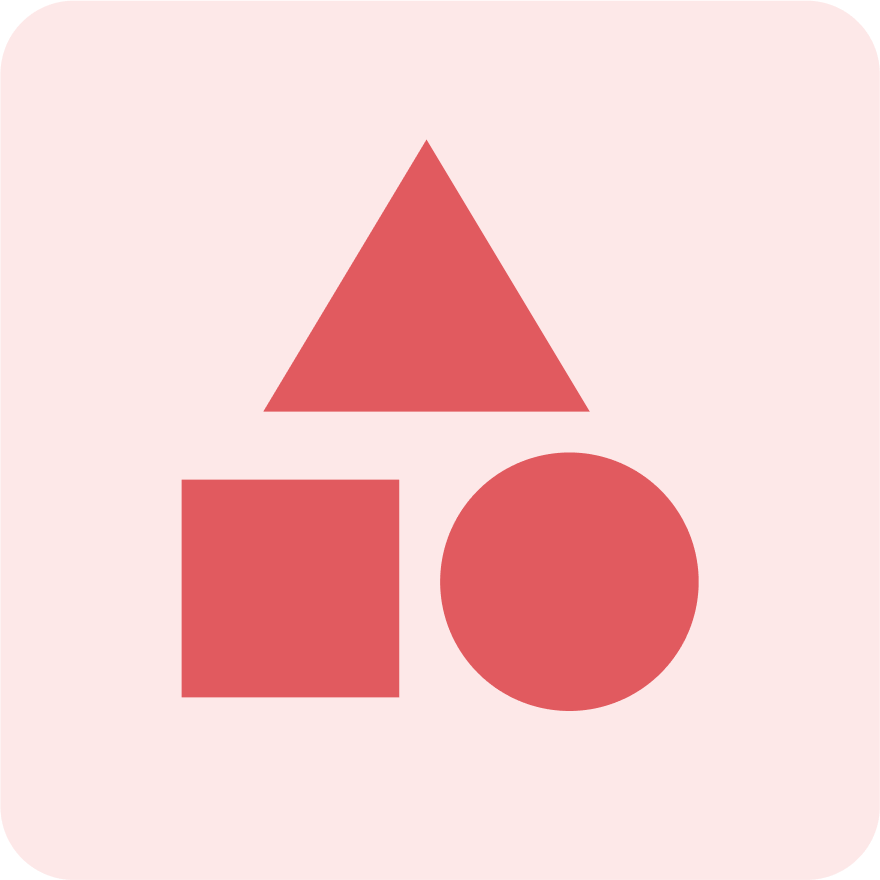}
\end{wrapfigure} 

Geometric shapes influence emotional responses by encoding affective and cognitive cues in their structural properties. Circular shapes are associated with positive valence and reduced cognitive load \cite{plass2014emotional}. Plass et al. \cite{plass2014emotional} found that circular designs increased positivity (PAS +4.0), reduced perceived task difficulty ($M = 2.80$ vs. $3.33$, $d = 0.36$), and improved comprehension ($M = 11.89$ vs. $10.45$, $d = 0.48$).  In UI design, circular dialog boxes attracted more early fixations and were preferred by a majority of Android (52\%) and iOS (45\%) users, enhancing perceived usability and emotional appeal \cite{yan2023case}.

Angular shapes (e.g., triangles, zigzags) signal sharpness and urgency, increasing arousal and drawing attention \cite{ferrara2015measuring, ames2009ll}. Luo et al. \cite{luo2021effects} found that long, angular taillights attracted more fixations at close range, underscoring their visual salience. In UI design,  angular forms are commonly used for warnings due to their perceived authority \cite{yan2023case}. Jagged line graphs similarly elicit greater arousal than smooth curves, likely due to their sharper contours and higher cognitive demand \cite{blair2024quantifying}.

Elongated linear shapes (e.g., wide rectangles) convey stability and direction. Compared to squares and short lines, they reduce fixation durations (895 ms vs. 927 ms) and accelerate visual search times (656 ms vs. 908 ms), especially at longer viewing distances \cite{luo2021effects}. These characteristics enhance navigability and structural clarity in interface design \cite{yan2023case}.

Irregular shapes, such as stars or fragmented forms, often evoke neutral to negative affect due to their asymmetry. Similarly, angular or pointed letterforms can disrupt reading fluency and evoke distinct emotional responses through their perceptual salience \cite{Medved2023influence}. Although their novelty may transiently increase arousal, such shapes are generally perceived as visually unstable and emotionally aversive \cite{blair2024quantifying}.

\subsubsection{Layout}
\begin{wrapfigure}{l}{0.06\textwidth}
  \vspace{-11pt} 
        \includegraphics[width=0.07\textwidth, trim=0 0 0 25, clip]{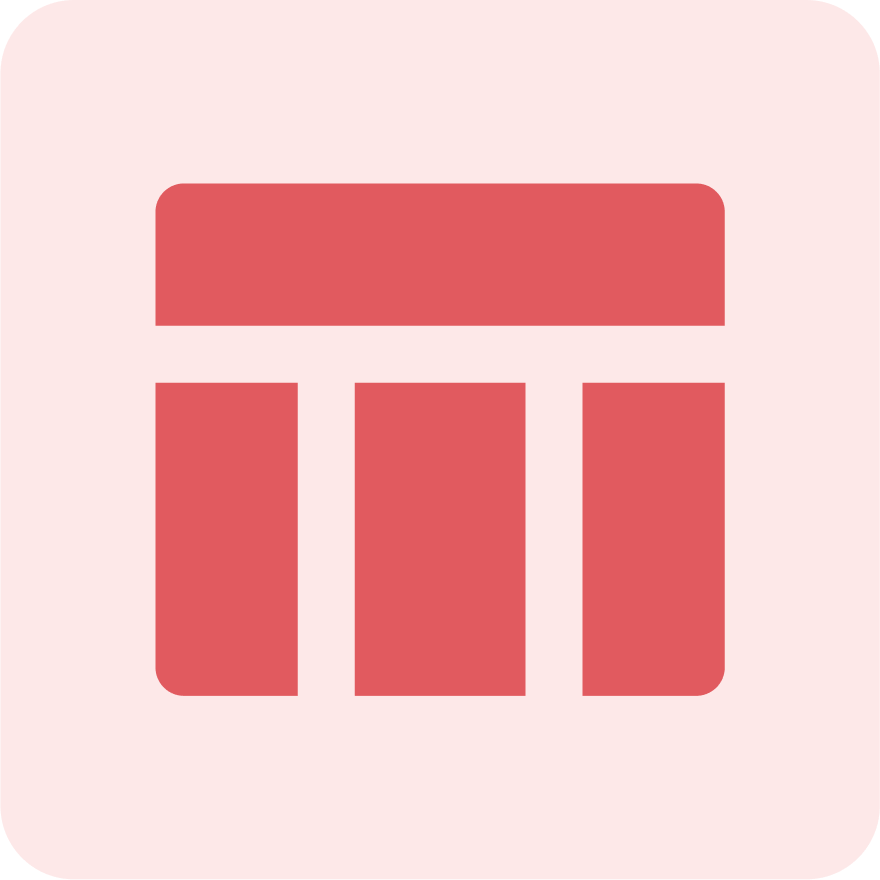}
\end{wrapfigure} 

Layout design shapes emotional and behavioral responses by organizing spatial structure and visual hierarchy. Empirical studies show that layouts following principles such as alignment and emphasis facilitate information processing and elicit positive affect \cite{mai2011rule, makin2012implicit, lu2017investigation}.

Layouts based on the rule of thirds—which position key elements along dividing lines or intersections—enhance visual appeal by guiding attention and reducing cognitive load, thereby increasing processing fluency and pleasure \cite{mai2011rule, Koliska2023guided}. 
Makin et al. \cite{makin2012implicit} found that symmetry elicited strong positive affective associations ($d = 1.04$, $p < 0.001$), indicating that symmetrical layouts promote comfort and fluency. Moderate asymmetry, such as diagonal alignment, enhances engagement by adding visual tension, whereas excessive asymmetry may disrupt focus and impair comprehension \cite{lu2017investigation}.

Circular layouts, including radial and looped layouts, are perceived as friendly and approachable due to their curvilinear forms. Lu et al. \cite{lu2017investigation} reported a weak but positive correlation between circularity and affective response ($r = 0.01$), potentially reflecting lower psychological defensiveness toward curved forms, Gómez-Puerto et al. \cite{GomezPuerto2016} showing that curved contours led to positive feelings and sharp transitions in contour triggered a negative bias. 

By contrast, complex layouts—characterized by high visual density or fragmentation—are associated with lower satisfaction. Lu et al. \cite{lu2017investigation} found a negative correlation between layout complexity and pleasure ($r = -0.11$, $p < 0.001$), likely due to elevated cognitive load and reduced retrieval efficiency. Hierarchical structures, such as clearly separated titles, subtitles, and content blocks, help mitigate these effects by reducing clutter and supporting navigation \cite{mai2011rule}.

\subsubsection{Imagery}
\begin{wrapfigure}{l}{0.06\textwidth}
  \vspace{-11pt} 
        \includegraphics[width=0.07\textwidth, trim=0 0 0 25, clip]{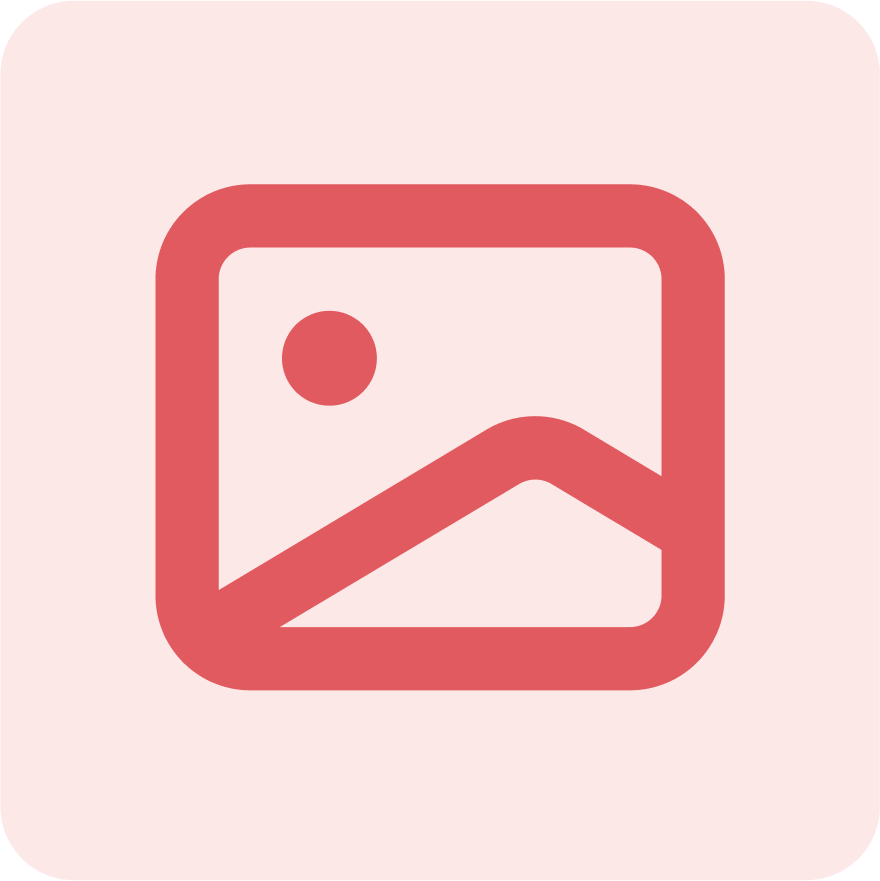}
\end{wrapfigure} 

Image design influences emotional experience and user behavior through visual features such as color, luminance, composition, and content \cite{lin2023effect, wang2017effects}.  

Images with warm hues, high saturation, and harmonious content—like natural landscapes or smiling faces—are typically associated with positive affect. These visual features enhance aesthetic appeal and support memory encoding \cite{lin2023effect, wang2017effects}. Neuroimaging studies show activation of the medial prefrontal cortex and nucleus accumbens, while such imagery also encourages approach-oriented behaviors like smiling or social engagement \cite{kragel2023mesocorticolimbic}.

In contrast, images featuring cool tones, low luminance, or stark contrast—especially when depicting distressing themes (e.g., disaster, violence, or social rejection)—evoke negative affect and high arousal. These visual cues increase perceptual salience and attentional capture, reflected in heightened activation of the amygdala and dorsolateral prefrontal cortex \cite{reisch2020negative}. While such stimuli enhance vigilance and sensory processing, excessive arousal—caused by extreme violence or overwhelming visual complexity—can tax cognitive resources, impair working memory, and hinder comprehension of core content \cite{lenski2023emotional}.



\subsection{Audio Design}  
\begin{figure*}[hbt!]
\setlength{\intextsep}{10pt plus 2pt minus 2pt}
    \centering
    \includegraphics[width=17.6cm]{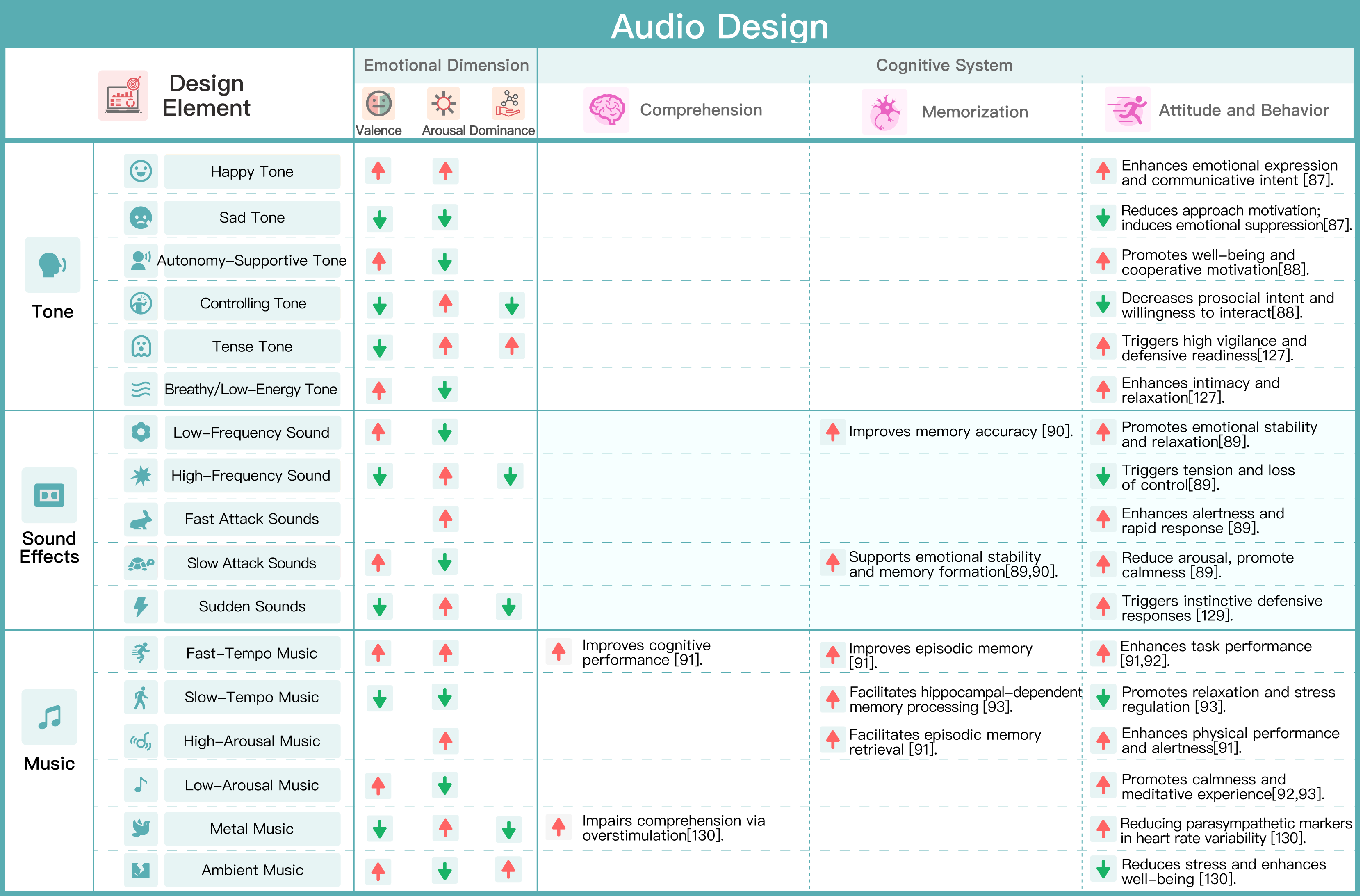}
    \caption{Emotional Dimensions and Cognitive-Behavioral Activation in Audio Design Elements.}
   \vspace{-4mm}
\label{fig:audio_design}
\end{figure*}

Audio design plays a critical role in shaping message reception and regulating user emotions, primarily through tone, sound effects, and music  (FIG.~\ref{fig:audio_design}).


\subsubsection{Tone}
\begin{wrapfigure}{l}{0.06\textwidth}
  \vspace{-11pt} 
        \includegraphics[width=0.07\textwidth, trim=0 0 0 25, clip]{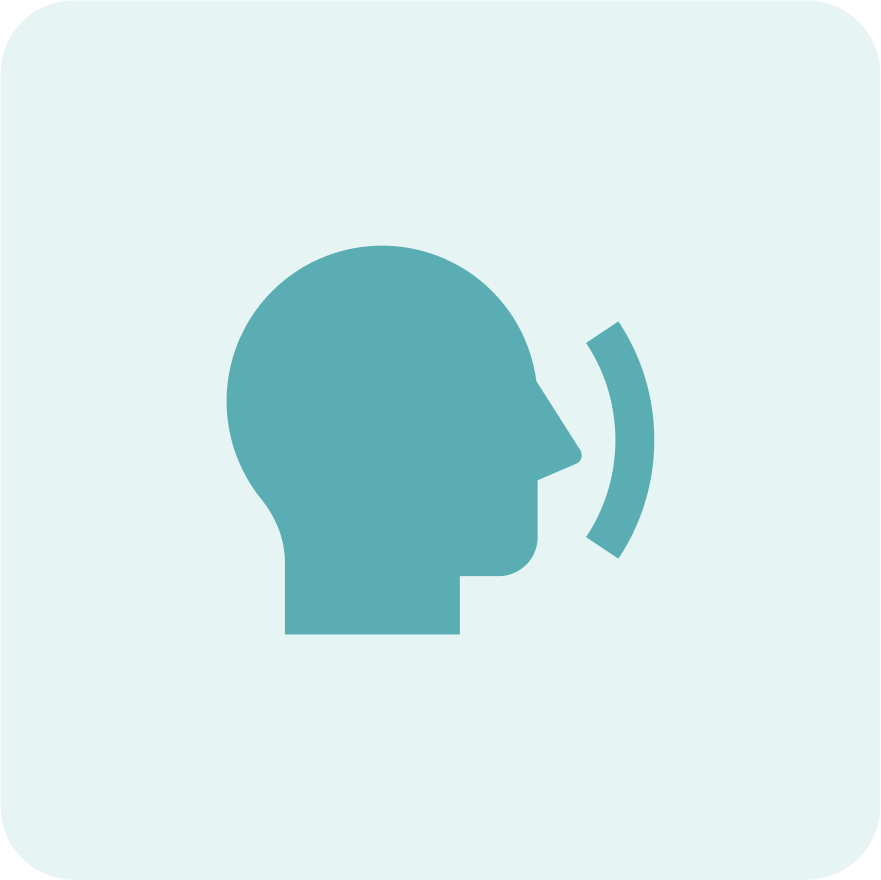}
\end{wrapfigure} 
Tone—defined as the speaker’s vocal quality, including pitch, intensity, and rhythm—modulates emotional communication via distinct neural and physiological mechanisms \cite{bestelmeyer2017effects}, shaping perceived valence \cite{weinstein2018you}, arousal \cite{gobl2003role}, and well-being \cite{schirmer2010mark}.

Positive tones, such as happy or autonomy-supportive tones, enhance emotional valence through neural pathways \cite{schirmer2010mark, weinstein2018you}. Schirmer \cite{schirmer2010mark} found that a happy tone increased perceived valence of neutral words ($M = 0.52$ vs. $0.35$; $F(1,30) = 4.89$, $p < .05$), associated with activation in the right inferior frontal gyrus and insula.
An autonomy-supportive tone, characterized by low vocal intensity and slower speech, boosted perceived autonomy ($\beta=0.13$, $p=0.02$), reduced stress ($\beta=-0.48$, $p<.001$), and enhanced well-being ($\beta=0.18$, $p<.001$) \cite{weinstein2018you}.

Negative tones, such as sad or controlling speech, reduce valence and increase psychological strain. Sad tones decreased the perceived valence of neutral words ($M = 0.23$ vs. $0.43$; $F(1,30) = 8.09$, $p < .01$), with activation in the left inferior frontal gyrus and amygdala \cite{schirmer2010mark}. Controlling tones, marked by high intensity and rapid delivery, increased stress ($\beta = 0.48$, $p < .001$) and reduced prosocial intent ($\beta = 0.45$, $p < .001$) \cite{weinstein2018you}. 

Different tone types also modulate arousal. Tense tones, with increased energy in the 0–5 kHz range, are perceived as angry and confident, and elicit activation in the amygdala and insula \cite{gobl2003role}. In contrast, breathy tones, marked by reduced energy in the 5–8 kHz range, are perceived as relaxing and intimate, and are linked to reduced limbic activity \cite{gobl2003role}. Arousal-inducing tones activate the bilateral superior temporal gyri (STG/STS; right peak: $F(2,50) = 50.07$, $p < .001$) \cite{weinstein2018you}, while tones with strong valence activate regions involved in emotional regulation, such as the prefrontal cortex and hippocampus \cite{bestelmeyer2017effects}. 

Additionally, sublexical acoustic features—such as specific vowel–consonant combinations—can shape emotional perception independently of meaning. Adelman et al. \cite{adelman2018emotional} showed that individual phonemes, particularly those at word-initial positions, modulate valence ratings (e.g., English \textipa{/\textturnv/} correlates with negativity, \textipa{/i:/} with positivity), suggesting that phonetic units convey affective cues at a pre-semantic level.

\subsubsection{Sound Effects}
\begin{wrapfigure}{l}{0.06\textwidth}
   \vspace{-11pt} 
        \includegraphics[width=0.07\textwidth, trim=0 0 0 25, clip]{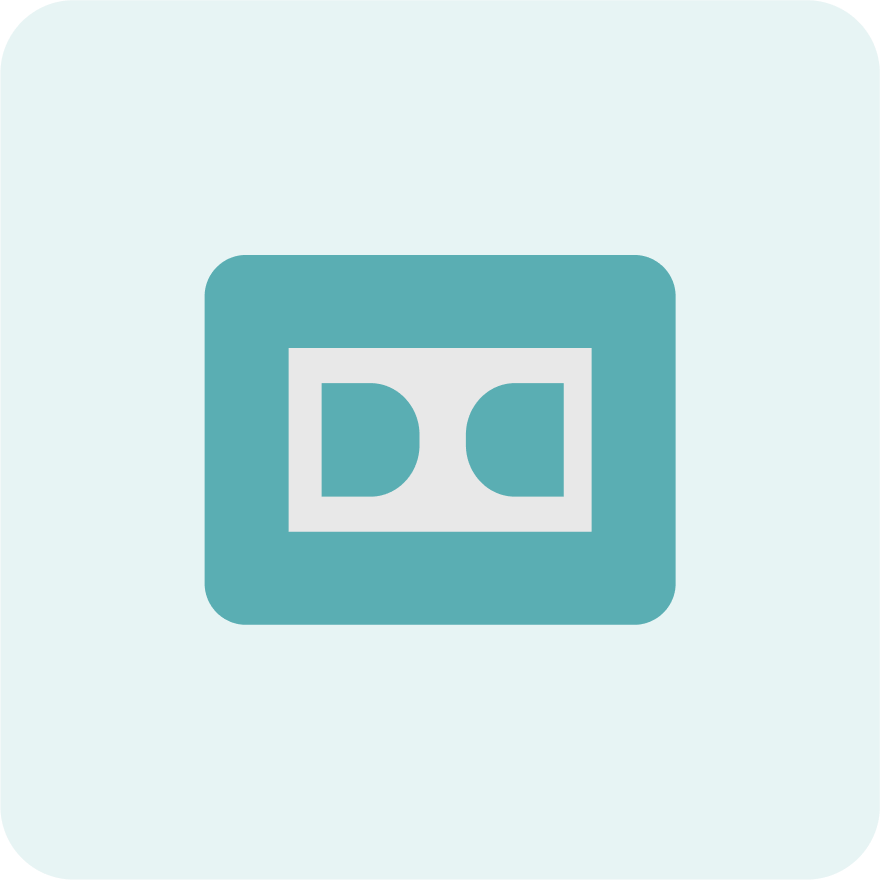}
\end{wrapfigure} 
Sound effects—non-verbal, non-musical sounds with distinct acoustic features—shape emotional responses through elements such as frequency content, attack slope, and spectral regularity \cite{eerola2012timbre}. 

Low-frequency sounds (e.g., organ tones, string instruments) are consistently associated with positive affect. Characterized by slow attack slopes, smooth envelopes, and stable spectra, they reduce arousal ($\beta = 0.28$, $p < .01$), enhance valence, and promote calmness \cite{eerola2012timbre}. Low-frequency-dominant auditory stimuli, particularly when precisely synchronized with the upstate phase of slow-wave oscillations (0.5-3 Hz) during sleep, significantly enhance memory consolidation. This memory-enhancing effect relies on the exact temporal alignment between auditory cues and slow-wave phases, facilitating neural reactivation between the hippocampus and neocortex \cite{batterink2016phase}. 

In contrast, high-frequency and spectrally irregular sounds (e.g., alarms, sharp electronic tones) are associated with negative emotions such as anxiety and fear. Characterized by rapid onsets and unstable spectra, they heighten physiological arousal, increasing heart rate and amygdala activity \cite{nardelli2015recognizing}. However, emotional interpretation is context-dependent. In musical settings, similar spectral features (e.g., synthesized high-frequency beats) can enhance perceived energy and motivation, whereas in non-musical contexts—such as emergency alerts—they consistently evoke tension \cite{eerola2012timbre}. Spectral content further shapes perception: sounds with low high-to-low frequency (HF–LF) ratios (e.g., organ tones) tend to induce calm, while those with high ratios (e.g., trumpet tones) are more likely to signal urgency or unease \cite{eerola2012timbre}. 


\subsubsection{Music}
\begin{wrapfigure}{l}{0.06\textwidth}
    \vspace{-11pt} 
        \includegraphics[width=0.07\textwidth, trim=0 0 0 25, clip]{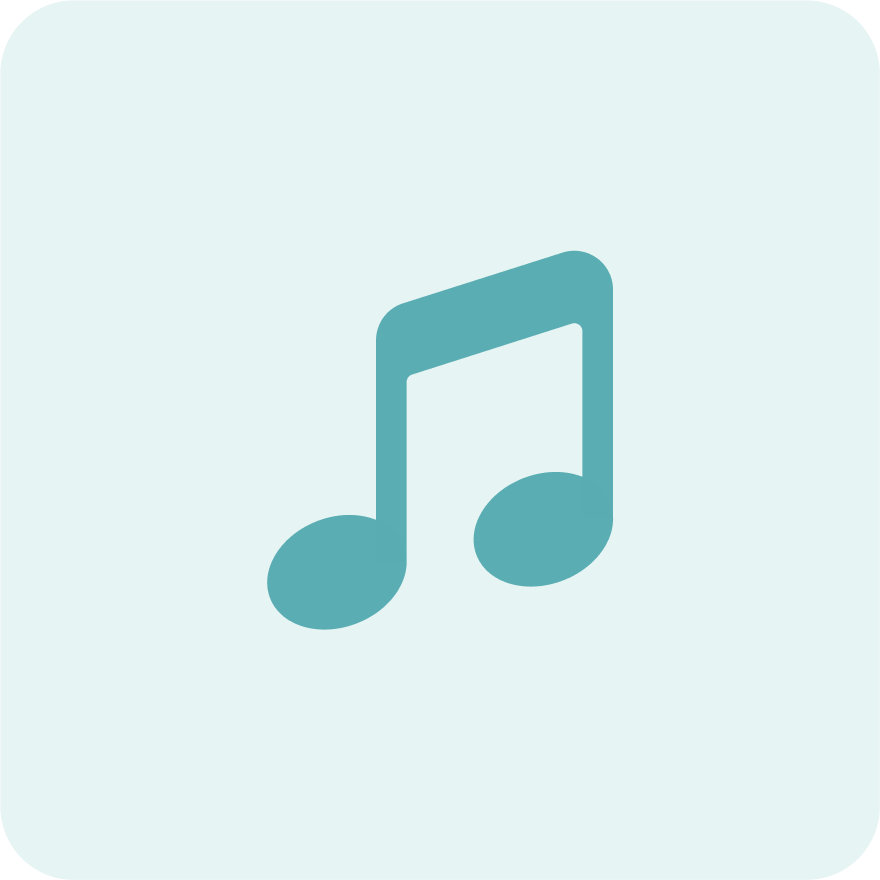}
\end{wrapfigure} 

Music modulates emotional and physiological states through its structural and acoustic features, such as tempo, mode (major/minor), harmony, and timbre \cite{bottiroli2014cognitive, salimpoor2011anatomically, koelsch2014brain}.

Specific types of music consistently evoke distinct emotional and cognitive responses. For example, fast-tempo ($\geq$ 130 BPM), major-key enhances arousal and improves cognitive performance, significantly boosting episodic memory (free recall: $F(3,192)=7.68$, $\eta^2 = 0.11$) and phonemic fluency (letter fluency task: $F(3,192)=9.70$, $\eta^2 = 0.13$) compared to silence or noise conditions \cite{bottiroli2014cognitive}. These effects are associated with dopaminergic activation in the nucleus accumbens during peak pleasure and caudate engagement during anticipation \cite{salimpoor2011anatomically}. Slow-tempo (60–90 BPM), minor-key music reduces physiological arousal and facilitates hippocampal-dependent memory processing \cite{koelsch2014brain}. Notably, both music types enhanced memory performance in older adults, suggesting emotion intensity rather than valence drives these benefits \cite{bottiroli2014cognitive}.

Music genre shapes emotional effects. Metal music, characterized by intensity and dissonance, elicits high arousal with negative valence, reducing parasympathetic markers (SDNN, RMSSD, HF) in heart rate variability. In contrast, ambient music (e.g., lullabies), marked by soft dynamics and minimal rhythmic variation, produces low arousal with positive valence, though it shows no significant impact on heart rate variability measures \cite{dimitriev2023effect}.

\subsection{Interaction Design}
\begin{figure*}[hbt!]
\setlength{\intextsep}{10pt plus 2pt minus 2pt}
    \centering
    \includegraphics[width=17.6cm]{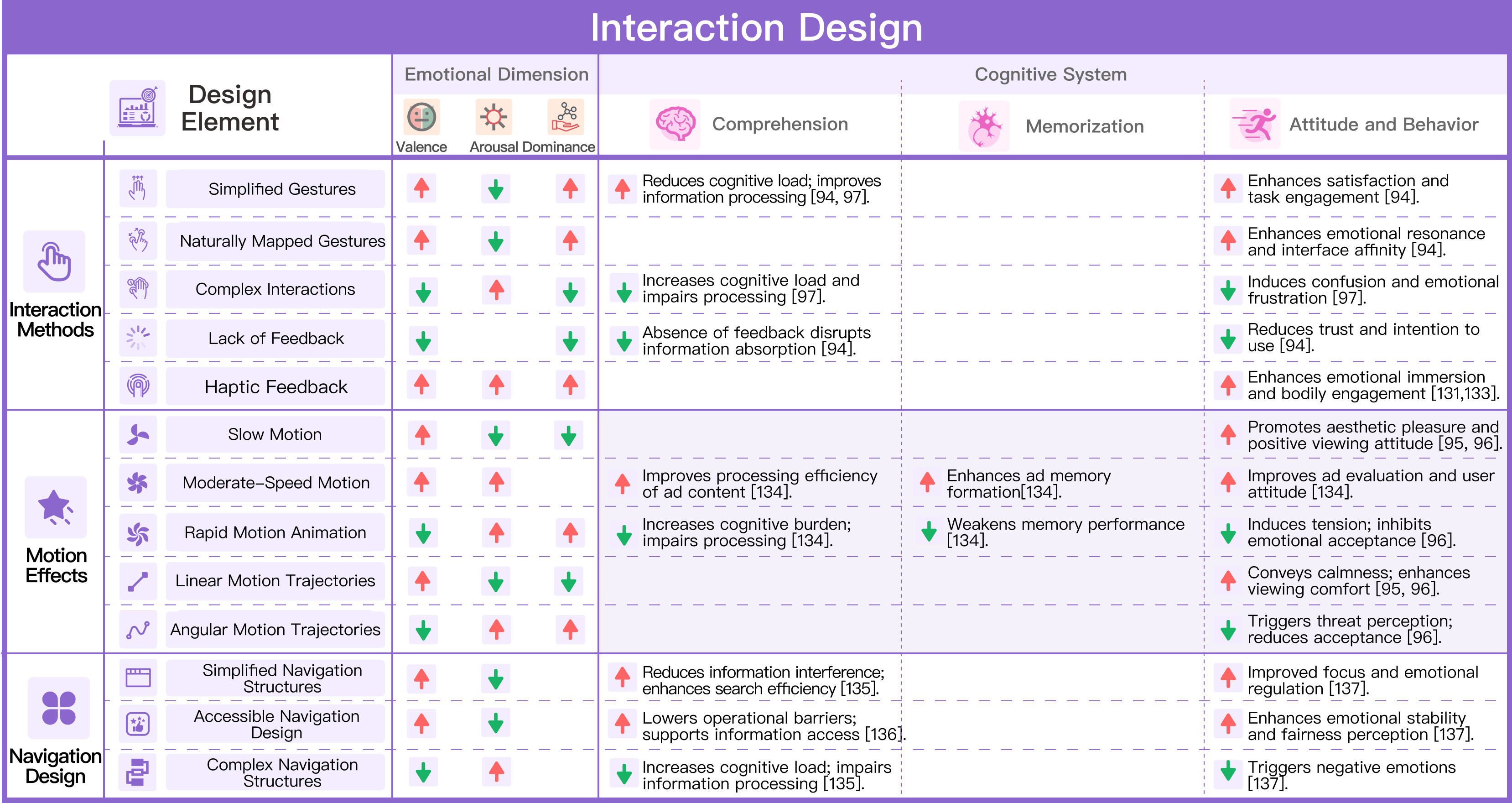}
    \caption{Emotional Dimensions and Cognitive-Behavioral Activation in Interaction Design Elements.}
   \vspace{-2mm}
\label{fig:interaction_design}
\end{figure*}

Interaction design seeks to optimize human–system interaction, enhancing both usability and emotional engagement. It involves key components such as interaction methods, motion effects, and navigation design (FIG.~\ref{fig:interaction_design}). 

 \subsubsection{Interaction Methods}
\begin{wrapfigure}{l}{0.06\textwidth}
        \includegraphics[width=0.07\textwidth, trim=0 0 0 25, clip]{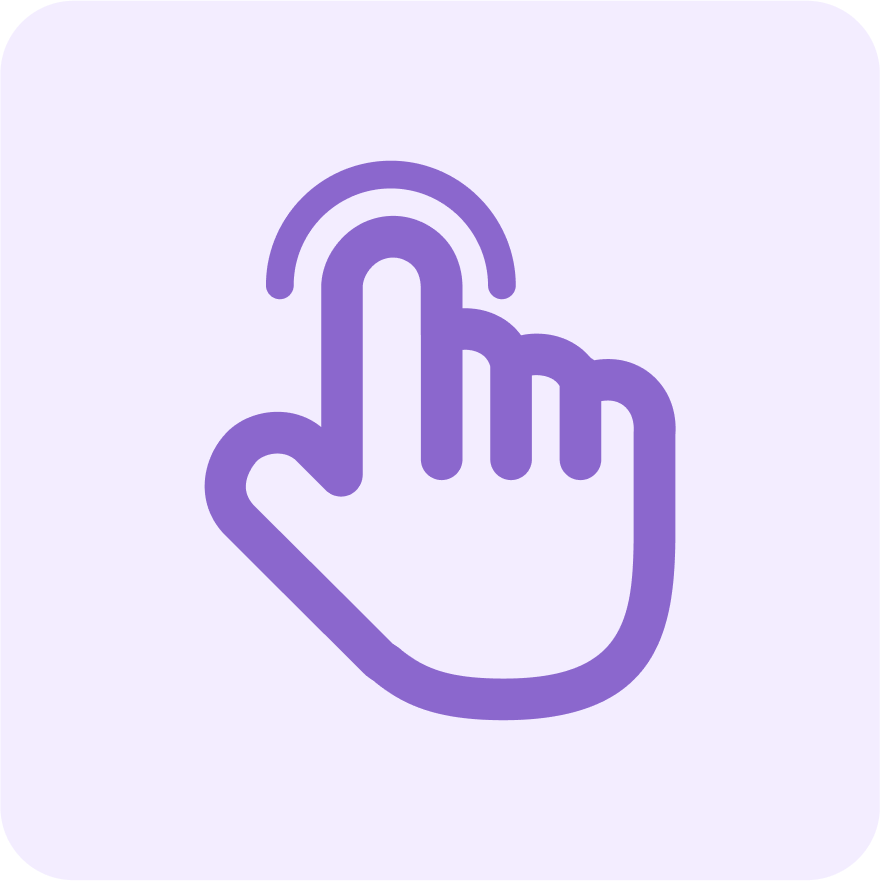}
\end{wrapfigure} 

Interaction methods—including input gestures and sensory feedback—determine how users engage with systems, thereby influencing emotional responses, cognitive effort, and perceived control \cite{sundar2014user, wodehouse2014exploring}.

Simplified gestures (e.g., tapping, swiping) enhance positive affect, user satisfaction, and engagement by lowering cognitive load and reinforcing emotional resonance \cite{sundar2014user, wodehouse2014exploring}. These intuitive interactions enhance perceived control and autonomy, particularly when aligned with natural motor schemas \cite{wodehouse2014exploring}. In contrast, complex gestures (e.g., 3D carousels) can elevate cognitive demand, often leading to confusion and frustration, which in turn reduce emotional valence and task efficiency \cite{sundar2014user}. Low perceived control, especially in systems lacking feedback, further diminishes satisfaction and elicits negative affect \cite{sundar2014user, wodehouse2014exploring}. 

Haptic feedback is particularly effective in enhancing emotional immersion, especially in remote or digital contexts where sensory input is limited. By incorporating tactile elements—such as keystroke resistance—and complementary auditory cues, it fosters bodily engagement and emotional satisfaction, counteracting the sensory flatness common in touchscreen-based interfaces \cite{wodehouse2014exploring}. Advanced haptic systems, such as vibrotactile gloves, further simulate realistic sensations that can trigger physiological responses (e.g., increased skin conductance) and activate limbic brain regions associated with emotional processing \cite{olugbade2023touch, saarinen2021social, sailer2022meaning}.  
 

\subsubsection{Motion Effects}
\begin{wrapfigure}{l}{0.06\textwidth}
  \vspace{-11pt} 
        \includegraphics[width=0.07\textwidth, trim=0 0 0 25, clip]{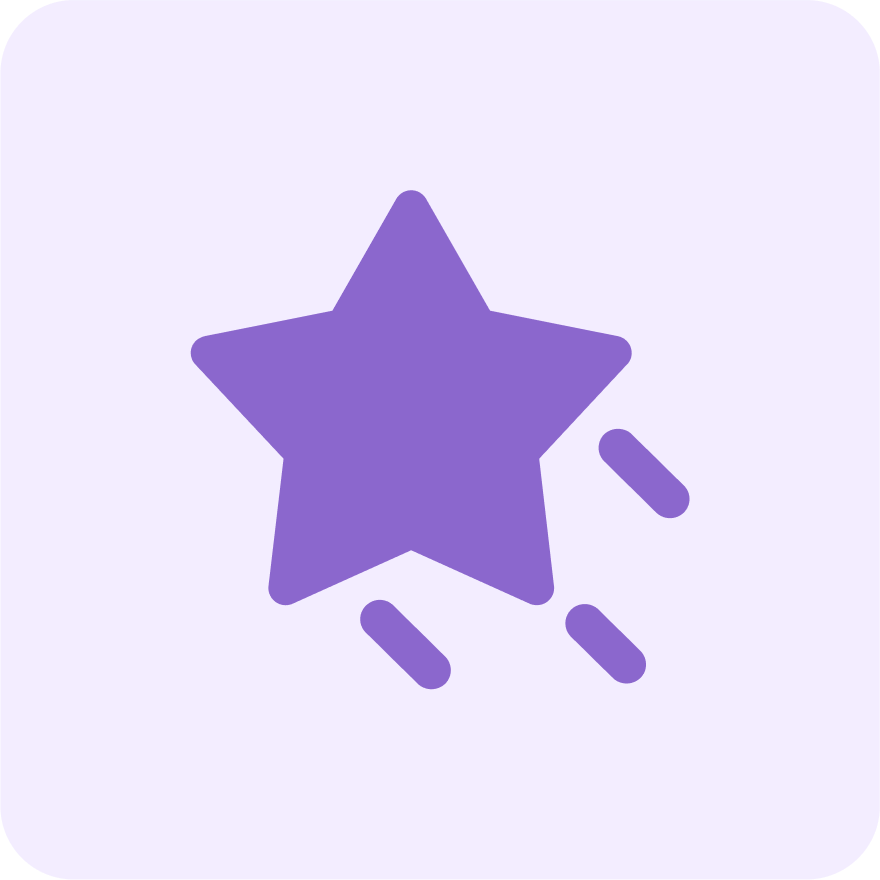}
\end{wrapfigure} 

Motion effects—defined as animated changes in visual elements over time, including parameters such as speed, trajectory, and frequency—play a crucial role in shaping users’ emotional and physiological responses \cite{lockyer2012affective, yoo2005processing}.

Slow motion consistently enhances positive affect \cite{wollner2018slow, lockyer2012affective, yoo2005processing}. Wöllner et al. \cite{wollner2018slow}  found that slow-motion video increased emotional valence compared to real-time playback ($F(1,39) = 10.65$, $p < .01$), with ballet scenes rated more positively than film or sports footage. Similarly, motion trajectory influences emotion: smooth, linear paths promote calmness and positive valence (e.g., $M = +0.82$), whereas sharp, angular movements evoke arousal, dominance, and perceived threat ($M = +1.05$ vs. $M = -0.96$, $p < .001$) \cite{lockyer2012affective}.

Motion speed further modulates emotional responses. Fast motion (12 pixels/second) increases urgency and perceived threat, while slower motion (4 pixels/second) fosters relaxation \cite{lockyer2012affective}. However, excessive or overly frequent animation—such as looping or abrupt transitions—can reduce emotional engagement by inducing hyperarousal and negative cognitive evaluations. High-frequency animations (more than once every 10 seconds) elevate arousal levels (5.05 vs. 3.98), lower cognitive valence ($-0.54$), and reduce attitudes toward content ($A_{ad} = 2.21$) \cite{lockyer2012affective, yoo2005processing}. 

Physiological evidence supports these findings. Real-time playback and angular trajectories increase respiratory rate and pupil dilation—markers of heightened arousal—while slow, linear motion reduces arousal markers and promotes relaxation \cite{wollner2018slow, lockyer2012affective}. Overall, smooth, slow motion fosters positive emotional states and mitigates physiological stress.

\subsubsection{Navigation Design}
\begin{wrapfigure}{l}{0.06\textwidth}
        \includegraphics[width=0.07\textwidth, trim=0 0 0 25, clip]{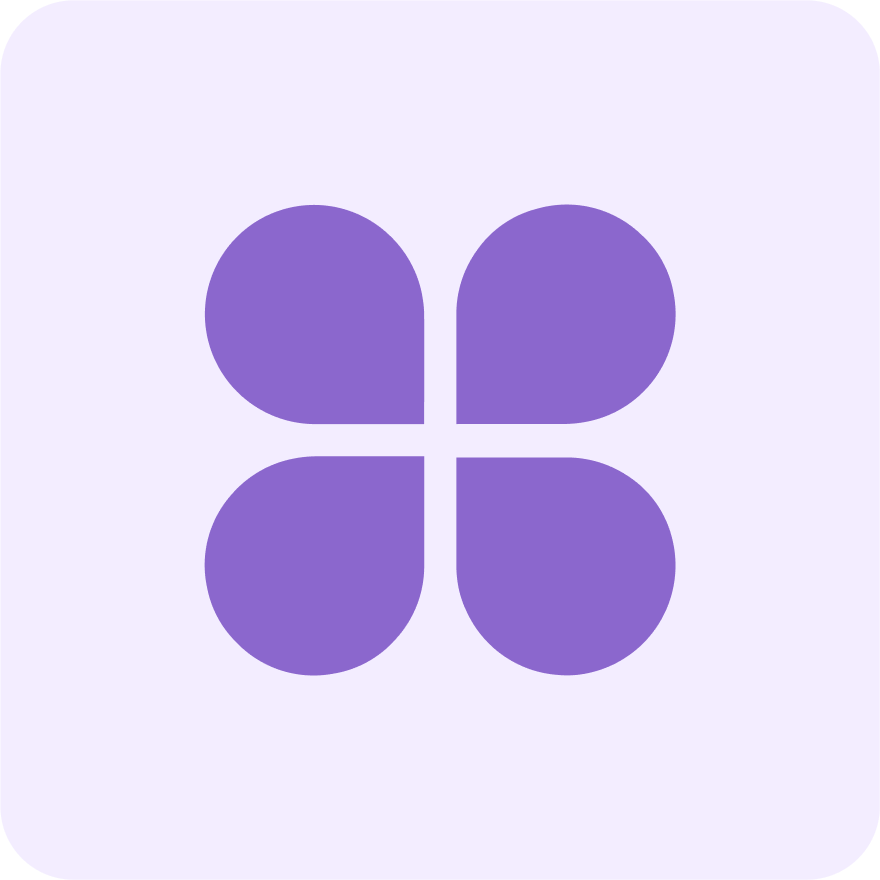}
\end{wrapfigure} 
Navigation design refers to how interface elements are structured to guide user interaction, including menus, links, and responsiveness \cite{abdelaal2023accessibility}.

Simplified navigation structures (reduced clutter, clear pathways) enhance positive affect by lowering visual complexity, which improves emotional valence and reduces stress markers like GSR. For visually impaired users, layered spatial models \cite{cai2024spatial} lead to higher task success rates and lower GSR—a finding mirrored in eye-tracking data showing consistent gaze patterns and stable arousal during structured tasks \cite{jin2018eye}.

Conversely, complex navigation (multi-level menus, unresponsive elements) elevates cognitive load and frustration. Sighted users show increased GSR during session timeouts, while visually impaired users exhibit sustained stress responses from inaccessible menus ($M = 3.06$ vs. $M = 0.93$, $p = .009$) \cite{abdelaal2023accessibility}. Poor design disrupts emotional balance: erratic gaze patterns and GSR peaks correlate with task failure \cite{jin2018eye}, whereas simplified layouts align with eye-tracking evidence of improved focus and emotional regulation.






\section{Summary and future directions} 
This review examined how emotional dimensions—valence, arousal, and dominance—differentially influence the three key stages of information processing: comprehension, memory, and behavior. Positive emotions generally broaden attention, support memory retention, and promote sharing, while negative emotions enhance analytic thinking and detail recall but may increase bias or reduce sharing, depending on context. High arousal boosts memory consolidation and sharing but risks spreading misinformation; low arousal supports reflective learning and may reduce engagement. High dominance fosters active participation, whereas low dominance leads to social withdrawal. These findings underscore the need to align emotional strategies with the cognitive and behavioral goals of information dissemination.

\textbf{Cross-domain Design Commonalities and Differences.} Although text, visuals, audio, and interaction operate through different modalities, they all serve to regulate emotion in ways that enhance perception, comprehension, and memory. For instance, in high-valence, high-arousal contexts, design strategies often prioritize immediate feedback and emotional intensity to stimulate user motivation and facilitate broader information dissemination. While each modality contributes to emotional modulation, they do so through distinct means. For example, text design evokes emotional depth through narrative structure and content; visual design guides attention and mood using color, shape, layout, and imagery; audio design influences affective states through tone, sound effects, and music; and interaction design fosters a sense of control and engagement through interaction methods, motion effects, and navigational cues. These complementary strengths underscore the importance of integrating multiple modalities to create emotionally coherent and cognitively engaging experiences—an approach crucial for enhancing the effectiveness of information communication across domains.

\textbf{The progressive relationship between emotional regulation and information transmission.}  In high-arousal contexts, users respond more positively to immediate feedback, as it enhances perceived control and supports action-oriented processing. Conversely, low-arousal environments benefit from minimalist design that fosters attentional focus and sustained cognitive engagement. Emotional regulation in communication is not just about managing affect—it also shapes user behavior and enhances message effectiveness. High-valence, high-arousal designs capture attention, stimulate motivation, and promote rapid information spread, while high-valence, low-arousal designs foster psychological stability and support deeper cognitive processing. Low-valence, high-arousal designs are effective in urgent scenarios, triggering quick responses, whereas low-valence, low-arousal designs encourage reflective engagement through restrained expression. Ultimately, the alignment of emotional activation with higher-order cognitive processes enhances both the depth of emotional resonance and the efficacy of information transfer.

\textbf{Opportunities For Future Research.} 
Despite advances in emotion-regulating design, several key areas warrant further investigation. One promising direction involves the integration of real-time emotional regulation technologies. By integrating artificial intelligence and biofeedback, design elements—such as text, visuals, sound, and interaction—can dynamically adapt to users’ emotional states, improving communication efficiency. Optimizing multimodal integration is another priority, particularly in aligning emotional cues across modalities, minimizing interference, and leveraging contrasts to sustain attention. Cultural adaptability also demands attention, as emotional preferences vary across cultural contexts. Developing emotionally resonant designs for global audiences remains a significant challenge. Finally, ethical considerations are essential. Emotion-driven design must balance communicative impact with user well-being, avoiding manipulation that risks psychological strain or emotional polarization. Addressing these issues will expand the responsible application of emotional design in communication, education, entertainment, and public services.


\bibliography{main}





\noindent\textbf{Acknowledgements}\\
We would like to thank Ruoyan Chen for her valuable contributions in creating the figures and assisting with other graphical elements of this paper. This work was supported by the National Key Research and Development Program of China (2023YFB3107100), and China Postdoctoral Science Foundation(2023M732674).\\

\noindent\textbf{Author contributions}\\
Shixiong Cao and Nan Cao contributed equally to this work. Shixiong Cao and Nan Cao conceived the study, conducted the literature review, synthesized the theoretical framework, and wrote the manuscript. Nan Cao supervised the project and served as the corresponding author.\\

\noindent\textbf{Competing interests}\\
The authors declare no competing interests.




\end{document}